\documentclass[prd,amsmath,amssymb,superscriptaddress,preprintnumbers,nofootinbib,twocolumn,10pt]{revtex4-1}

\pdfoutput=1

\usepackage{graphicx}
\usepackage{dcolumn}
\usepackage{booktabs}
\usepackage{multirow}
\usepackage{threeparttable}
\newcommand{\ve}{\varepsilon}
\usepackage[colorlinks=true, linkcolor=blue, citecolor=blue]{hyperref}

\begin{document}

\title{Multi-messenger standard-siren cosmology for third-generation gravitational-wave detectors: forecasts considering observations of gamma-ray bursts and kilonovae}

\author{Tao Han}
\affiliation{Liaoning Key Laboratory of Cosmology and Astrophysics, College of Sciences, Northeastern University, Shenyang 110819, China}

\author{Jing-Fei Zhang}
\thanks{jfzhang@neu.edu.cn}
\affiliation{Liaoning Key Laboratory of Cosmology and Astrophysics, College of Sciences, Northeastern University, Shenyang 110819, China}

\author{Xin Zhang}%\footnote{Corresponding author}}
\thanks{zhangxin@neu.edu.cn (corresponding author)}
\affiliation{Liaoning Key Laboratory of Cosmology and Astrophysics, College of Sciences, Northeastern University, Shenyang 110819, China}
\affiliation{MOE Key Laboratory of Data Analytics and Optimization for Smart Industry, Northeastern University, Shenyang 110819, China}
\affiliation{National Frontiers Science Center for Industrial Intelligence and Systems Optimization, Northeastern University, Shenyang 110819, China}

\begin{abstract}
	
In the third-generation (3G) gravitational-wave (GW) detector era, GW multi-messenger observations for binary neutron star merger events can exert significant effects on exploring the cosmic expansion history. Extending a previous work, we explore the potential of 3G GW standard siren observations in cosmological parameter estimation by considering their associated electromagnetic (EM) counterparts, including $\gamma$-ray burst (GRB) coincidence observations by the Gravitational Wave High-energy Electromagnetic Counterpart All-sky Monitor and GW-triggered target-of-opportunity observations of kilonovae by different optical survey projects. During an assumed 10-year observation, we predict that the number of detectable GW-kilonova events is $\sim 4900$ with redshifts below $\sim 0.4$ under the GW detector network and Large Synoptic Survey Telescope in the $i$ band, which is more than three times that of GW-GRB detections. For the cosmological analysis, we find that with the inclusion of GW-kilonova detections, the constraints on cosmological parameters from GW-EM detections are significantly improved compared to those from GW-GRB detections. In particular, GW-EM detections can tightly constrain the Hubble constant with precision ranging from $0.076\%$ to $0.034\%$. Moreover, GW multi-messenger observations can effectively break the cosmological parameter degeneracies generated by the typical EM observations, CMB+BAO+SN (CBS). The combination of CBS and GW-EM can tightly constrain the equation-of-state parameters of dark energy $w$ in the $w$CDM model and $w_0$ in the $w_0w_a$CDM model with precision of $0.72\%$ and $0.99\%$, respectively, meeting the standard of precision cosmology. In conclusion, GW multi-messenger observations could play a crucial role in helping solve the Hubble tension and probing the fundamental nature of dark energy.

\end{abstract}

\maketitle

\section{Introduction}

The precise measurements of cosmic microwave background (CMB) anisotropies by the Planck mission have ushered in a new era of precision cosmology~\cite{Planck:2018vyg}. Based on precise measurements, the $\Lambda$ cold dark matter ($\Lambda$CDM) model, widely regarded as the standard model of cosmology, has been strongly favored, with dark energy being described by the cosmological constant $\Lambda$. However, recent precise measurements of cosmological parameters have revealed several puzzling tensions. This confusion arises from a more-than-5$\sigma$ discrepancy between the values of the Hubble constant ($H_0$) inferred from the CMB observation (assuming the $\Lambda$CDM model)~\cite{Planck:2018vyg} and those obtained through the model-independent distance ladder measurement~\cite{Riess:2021jrx}, which is now commonly regarded as a significant crisis for cosmology~\cite{Riess:2019qba,Verde:2019ivm}. The Hubble tension has been intensively discussed in the literature~\cite{Li:2013dha,Zhang:2014nta,Zhao:2017urm,cai:2020,Guo:2018ans,Guo:2019dui,Yang:2018euj,Vagnozzi:2019ezj,DiValentino:2019jae,DiValentino:2019ffd,Feng:2019jqa,Lin:2020jcb,Li:2020tds,Hryczuk:2020jhi,Gao:2021xnk,Gao:2022ahg,Cai:2021wgv,Vagnozzi:2021tjv,Vagnozzi:2021gjh,Li:2024qso,Li:2024qus,Li:2025owk,Poulin:2018cxd,Schoneberg:2021qvd}. Possible explanations for this tension include not only new physics beyond the standard $\Lambda$CDM model, but also potential systematic uncertainties in the distance-ladder calibration (e.g., Cepheid measurements)~\cite{Riess:2021jrx}, in the CMB and baryon acoustic oscillation (BAO) analyses~\cite{Planck:2018vyg}, or changes in the sound speed at different epochs~\cite{Poulin:2018cxd,Schoneberg:2021qvd}. At present, no consensus has been reached on a definitive resolution. Therefore, it is crucial to explore independent cosmological probes that can measure $H_0$ with high precision and offer complementary insights into this tension. In this context, the gravitational-wave (GW) standard siren method provides a promising and independent approach to probing cosmology.

The discoveries of GW signals generated by the inspiral and merger of compact binary systems mark the beginning of the era of GW astronomy~\cite{LIGOScientific:2016aoc}. GW can be used as a standard siren to explore the evolution of the universe, since the luminosity distance to the source can be obtained through the analysis of the GW waveform without relying on the cosmic distance ladder~\cite{Schutz:1986gp,Holz:2005df}. If the redshift of the source can also be obtained, then we can establish the true distance-redshift relation to measure various cosmological parameters and explore the expansion history of the universe (see, e.g., Refs.~\cite{Zhao:2010sz,Cai:2017aea,Cai:2016sby,Wang:2018lun,Zhang:2019ylr,Zhang:2019loq,Zhang:2018byx,Li:2019ajo,Jin:2020hmc,Zhang:2019ple,Jin:2023zhi,Jin:2021pcv,Wu:2022dgy,Jin:2022tdf,Jin:2023tou,Wang:2019tto,Wang:2021srv,Zhao:2019gyk,Jin:2023sfc,Jin:2025dvf,Han:2024sxm,Jin:2022qnj,Han:2023exn,Belgacem:2019tbw,Hou:2022rvk,Li:2023gtu,Qi:2021iic,Wang:2022oou,Xiao:2024nmi,Dong:2024bvw,Dong:2025ikq,Feng:2024lzh,Feng:2024mfx,Feng:2025wbz,Song:2022siz,Song:2025ddm,Zhang:2025yhi,KAGRA:2013rdx,2022SCPMA..6510431G} for related discussions).

The measurement of redshift is an important task for GW standard sirens. A primary method involves observing their electromagnetic (EM) counterparts, commonly referred to as bright sirens~\cite{Schutz:1986gp,Holz:2005df}. In cases where no counterparts are identified, statistical techniques are employed to infer their redshifts, referred to as dark sirens~\cite{MacLeod:2007jd}. Binary neutron star (BNS) mergers have long been thought to be the origin of short $\gamma$-ray bursts (GRBs)~\cite{Paczynski:1986px,Eichler:1989ve,Narayan:1992iy,Rosswog:2013wga} and kilonovae~\cite{Li:1998bw,Metzger:2010sy,Tanvir:2013pia,Metzger:2016pju}. Therefore, on the EM side, short GRBs  and kilonovae are typically considered as two types of counterparts to BNS mergers. On 17 August 2017, the first observed BNS merger GW event, i.e., GW170817, was detected by the LIGO-Virgo-KAGRA Collaboration~\cite{LIGOScientific:2017vwq}. This GW event was subsequently confirmed to be associated with a short GRB (GRB170817A)~\cite{LIGOScientific:2017zic,Goldstein:2017mmi,Savchenko:2017ffs,Zhang:2017lpb} and an ultraviolet-optical-infrared kilonova (AT2017gfo)~\cite{LIGOScientific:2017ync,Andreoni:2017ppd,Arcavi:2017xiz,Chornock:2017sdf,Coulter:2017wya,Covino:2017bjc,TOROS:2017pqe,Drout:2017ijr,Evans:2017mmy,Hu:2017tlb,Kilpatrick:2017mhz,Lipunov:2017dwd,McCully:2017lgx,Nicholl:2017ahq,Pian:2017gtc,Shappee:2017zly,Smartt:2017fuw,DES:2017kbs,Tanvir:2017pws,J-GEM:2017tyx,Valenti:2017ngx,Cowperthwaite:2017dyu,Kasliwal:2017ngb,Villar:2017wcc}. The redshift of GW170817, combined with the luminosity distance, led to a measurement of $H_0=70^{+12}_{-8}~\rm km~s^{-1}~Mpc^{-1}$, with precision of $\sim 14\%$~\cite{LIGOScientific:2017adf,Guidorzi:2017ogy}. However, except for GW170817, no other GW events associated with EM counterparts have been detected. This is because, according to the third observing run (O3) of the LIGO-Virgo-KAGRA Collaboration, the observed BNS merger rate ranges from 10 to 1700~$\rm Gpc^{-3}~yr^{-1}$~\cite{KAGRA:2021duu}, suggesting that nearby events similar to GW170817 may occur approximately once per decade, while the EM emissions from similar sources at greater distances would appear significantly fainter. As a result, the current measurements are far from achieving arbitration for the Hubble tension. Additional detectable GW multi-messenger events are needed to reduce the statistical uncertainty.

To date, GW multi-messenger observations remain rare. Even with forthcoming upgrades for the fifth observing run (O5, planned for 2028), the second-generation GW detectors are expected to reach redshifts of only $\sim 0.1$ for BNS mergers~\cite{O5-web}. On the EM side, short GRB detections are limited by jet beaming, while kilonova searches are challenged by their modest peak luminosity and rapid evolution compared to supernovae. These considerations motivate next-generation upgrades to both GW and EM facilities.

In the 2030s, the third-generation (3G) ground-based GW detectors, namely the Einstein Telescope (ET)~\cite{ET-web,Punturo:2010zz} and the Cosmic Explorer (CE)~\cite{CE-web,Evans:2016mbw}, are expected to become operational. ET is designed as an underground observatory with three low- and three high-frequency interferometers nested in a triangular shape, each arm being 10 km in length, while CE is an L-shaped surface interferometer with 40 km arms. With sensitivities exceeding those of current detectors by more than an order of magnitude~\cite{Evans:2021gyd}, these facilities will represent a substantial leap forward compared to the current instruments for the cosmological census of GW sources. Such capabilities will enable the detection of a substantially larger number of BNS mergers, reaching redshifts up to $\sim 5$, well above the star formation peak~\cite{Jin:2022qnj,Han:2023exn}. When operating in synergy with wide-field-of-view (FoV) GRB detectors such as the Gravitational Wave High-energy Electromagnetic Counterpart All-sky Monitor (GECAM)~\cite{LI:2020lsh}, this combination can enhance the ability for joint GW-GRB detections at significantly larger distances. Furthermore, the 3G GW detectors are expected to provide improved localization constraints for BNS GW events, especially when operating as a network, with a few tens of percent of detected BNS localized within $10~\rm{deg}^2$ and almost all within $100~\rm{deg}^2$~\cite{Li:2021mbo,Borhanian:2022czq,Mills:2017urp}. In conjunction with this improved localization, the commissioning of more sensitive optical survey projects, such as the Wide Field Survey Telescope (WFST)~\cite{2018AcASn..59...22S}, the Large Synoptic Survey Telescope (LSST; also known as the Vera Rubin Observatory)~\cite{LSSTScience:2009jmu}, and the Chinese Space Station Survey Telescope (CSST)~\cite{Gong:2019yxt}, will facilitate the search for associated kilonovae through target-of-opportunity (ToO) follow-up observations of 3G GW triggers.

In our previous work~\cite{Han:2023exn}, we investigated the potential of GW standard siren observations for cosmological parameter estimation, considering joint observations by 3G GW and future GRB detectors. However, for low-redshift BNS mergers (e.g., redshifts below $\sim 0.4$ with LSST in our simulations), kilonovae, which are of much lower luminosity than GRBs but have significantly larger observable angles, will become the primary targets for EM counterpart searches. In this paper, we extend our analysis of GW multi-messenger observations. We conduct a comprehensive analysis of GW multi-messenger observations using different EM detection scenarios, including GW-GRB, GW-kilonova, and their combined GW-EM observations, to explore their potential for cosmological parameter estimation.

To maximize the scientific return of 3G GW detectors on cosmological parameter estimation in a multi-messenger context, it is crucial to evaluate the expected joint GW-EM detections and determine the instrumental science requirements and the optimal detection strategy. In this work, we focus on the synergy of 3G GW detectors with wide-FoV GRB detector and different optical survey projects. For the GRB observations, we adopt GECAM, which has an all-sky FoV, a high sensitivity, and a wide energy interval. For the GW-triggered ToO observations of kilonovae, the survey projects we considered in this work include WFST, LSST, and CSST. For the cosmological analysis, we constrain three typical cosmological models, including the $\Lambda$CDM, $w$CDM, and $w_0w_a$CDM models. Here we highlight two points upgraded in this paper: (i) Distinct from previous work on this topic, we not only emphasize the measurement precision of cosmological parameters with a given number of joint detections, but also consider the practical limitations of EM detections using planned ground- and space-based telescope facilities. (ii) We present the first comprehensive analysis of GW multi-messenger detections for cosmological parameter estimation in different EM detection scenarios, including GW-GRB, GW-kilonova, and their combined GW-EM detections, which could better show their potential for cosmological parameter estimation in the 3G GW detector era.

This paper is organized as follows. In Sec.~\ref{sec2}, we overview the method for simulating GW signals and their associated EM counterparts. In Sec.~\ref{sec3}, we report the predicted results of the multi-messenger detections of GWs and EM counterparts. In Sec.~\ref{sec4}, we present the constraint results for cosmological parameters from the GW multi-messenger detections. The conclusion is given in Sec.~\ref{sec5}. Throughout this paper, the fiducial values of cosmological parameters are set to the constraint results from CMB (Planck 2018 distance priors), BAO, and type Ia supernova (SN) data. Unless otherwise specified, we set $G=c=1$.

\section{Methodology}\label{sec2}

\subsection{Simulation of BNS mergers}

To construct a catalog of BNS coalescences, we first produce the redshift distribution of BNS mergers over the redshift range from 0 to 10, which fully covers the detection horizon of 3G GW detectors~\cite{Belgacem:2019tbw,Han:2023exn}. It can be expressed as a normalized probability distribution
\begin{equation}
	p(z)=\frac{R_{\rm m}(z)}{\int_{0}^{10}R_{\rm m}(z){\rm d}z},
\end{equation}
where $R_{\rm m}(z)$ denotes the BNS merger rate as a function of redshift $z$ in the observer frame. It can be written as
\begin{equation}
	R_{\rm m}(z)=\frac{\mathcal{R}_{\rm m}(z)}{1+z} \frac{{\rm d}V(z)}{{\rm d}z},
\end{equation}
where ${\rm d}V/{\rm d}z$ is the comoving volume element and $\mathcal{R}_{\rm m}(z)$ is the BNS merger rate in the source frame.

A BNS merger can be thought of as occurring with a delay timescale relative to the BNS formation history, which can be written as
\begin{equation}
	\mathcal{R}_{\rm m}(z)=\int_{t_{\rm min}}^{t_{\rm max}} \mathcal{R}_{\rm f}[t(z)-t_{\rm d}] P(t_{\rm d}){\rm d}t_{\rm d},
\end{equation}
where $P(t_{\rm d})$ is the delay time distribution between the formation of the BNS system and the two-NS merger, $t_{\rm d}$ is the time delay between the formation of the BNS system and merger, $t(z)$ is the age of the universe at the time of merger, $t_{\rm min}=20~\rm Myr$ is the minimum delay time, $t_{\rm max}=t_{\rm H}$ represents a maximum delay time which is equal to the Hubble time~\cite{Belgacem:2019tbw}, and $\mathcal{R}_{\rm f}$ is the BNS formation rate. We assume that $\mathcal{R}_{\rm f}$ is proportional to the star formation rate (SFR) density:
\begin{equation}
	\mathcal{R}_{\rm f}\equiv \lambda \psi_{\rm MD},
\end{equation}
where $\psi_{\rm MD}$ is the Madau-Dickinson SFR~\cite{Madau:2014bja} and $\lambda$ is the BNS mass efficiency determined by the local comoving merger rate $\mathcal{R}_{\rm m}(z=0)$.

The main types of delay time distributions include the power-law delay model~\cite{Virgili:2009ca,DAvanzo:2014urr}, the exponential time delay model~\cite{Sathyaprakash:2019rom,Vitale:2018yhm}, the Gaussian delay model~\cite{Virgili:2009ca}, and the log-normal delay model~\cite{Nakar:2005bs,Wanderman:2014eza}. For simplicity, we only adopt the power-law delay model as our delay time model, which is given by $P(t_{\rm d})=1/t_{\rm d}$, with $t_{\rm d}>t_{\rm min}$.

In this paper, we adopt a local comoving merger rate of $105.5~\rm Gpc^{-3}~yr^{-1}$ estimated from the latest O3 observing run under a simplified population assumption~\cite{KAGRA:2021duu}, which lies within the range from 10 to 1700~$\rm Gpc^{-3}~yr^{-1}$mentioned above, as inferred from various population models. We simulate a catalog of BNS mergers for 10 years. For each source, the location $(\theta,\phi)$, the cosine of the inclination angle $\iota$, the polarization angle $\psi$, and the coalescence phase $\psi_{\rm c}$ are randomly drawn from uniform distributions. For the component masses in a BNS system, we assume that each follows an independent uniform distribution in the range $1.2~M_{\odot}$ to $2~M_{\odot}$~\cite{KAGRA:2021duu}.

\subsection{Detection of GW events}

In this section, we will briefly introduce the detection of a GW detector network. We use the vector $r_k$ with $k=1,2,\ldots,N$ to denote the spatial locations of the detectors, which is expressed as
\begin{equation}
\boldsymbol{r}_k=R_{\oplus}(\sin\varphi_k\cos\alpha_k,\sin\varphi_k\sin\alpha_k,\cos\varphi_k),
\end{equation}
where $R_{\oplus}$ is the radius of the Earth, and $\varphi_k$ is the latitude of the detector in the celestial system. We define $\alpha_k$ as $\alpha_k\equiv\lambda_k+\Omega_{\rm r}t$, where $\lambda_k$ is the east longitude of the detector and $\Omega_{\rm r}$ is the rotational angular velocity of the Earth. In this paper, we take the zero Greenwich sidereal time as $t=0$.

With the stationary phase approximation (SPA), the GW waveform of a detector network including $N$ independent detectors can be written as~\cite{Wen:2010cr}
\begin{equation}
\tilde{\boldsymbol{h}}(f)={\rm e}^{-{\rm i}\boldsymbol\Phi}\boldsymbol h(f),
\label{eq:waveform}
\end{equation}
where $\boldsymbol\Phi$ is the $N\times N$ diagonal matrix, with $\Phi_{ij}=2\pi f\delta_{ij}(\boldsymbol{n\cdot r}_i(f))$, and
\begin{widetext}
\begin{equation}
\begin{aligned}
\boldsymbol h(f)=&\left [\frac{h_1(f)}{\sqrt{S_{\rm {n},1}(f)}}, \frac{h_2(f)}{\sqrt{S_{\rm {n},2}(f)}}, \ldots,\frac{h_k(f)}{\sqrt{S_{{\rm n},k}(f)}},\ldots,\frac{h_N(f)}{\sqrt{S_{{\rm n},N}(f)}}\right ]^{\rm T},
\end{aligned}
\end{equation}
\end{widetext}
where $h_k(f)$ is the frequency domain GW waveform, and $S_{{\rm n},k}(f)$ is the one-side noise power spectral density of the $k$-th detector. In this work, we consider the waveform in the inspiraling stage for a BNS system and neglect the NS spins, which has been shown to have a negligible impact on the GW waveform of BNS systems~\cite{OShaughnessy:2006uzj}. We adopt the restricted post-Newtonian approximation and obtain the waveform to the 3.5 PN order. The SPA Fourier transform of the GW waveform of the $k$-th detector can be written as~\cite{Sathyaprakash:2009xs}
\begin{align}
	h_k(f)=&\mathcal A_k f^{-7/6}\exp
	\{{\rm i}[2\pi f t_{\rm c}-\pi/4-2\psi_c+2\Psi(f/2)]\nonumber\\ &-\varphi_{k,(2,0)}\},
\end{align}
where the Fourier amplitude $\mathcal A_k$ is given by
\begin{align}
	\mathcal A_k=&\frac{1}{d_{\rm L}}\sqrt{(F_{+,k}(1+\cos^{2}\iota))^{2}+(2F_{\times,k}\cos\iota)^{2}}\nonumber\\ &\times\sqrt{5\pi/96}\pi^{-7/6}\mathcal M^{5/6}_{\rm chirp},
\end{align}
where $d_{\rm L}$ is the luminosity distance to the source, $\mathcal M_{\rm chirp}$ is the chirp mass of the binary system, and the detailed forms of $\psi(f/2)$ and $\varphi_{k,(2,0)}$ can be found in Refs.~\cite{Zhao:2017cbb,Blanchet:2004bb}.
With SPA, $F_{+,k}$, $F_{\times,k}$, and $\Phi_{ij}$ are all functions of frequency, which can be written as
\begin{equation}
\begin{aligned}
	&F_{+,k}(f)=F_{+,k}(t=t_{\rm f}), ~~~~F_{\times,k}(f)=F_{\times,k}(t=t_{\rm f}),\\
	&\Phi_{ij}(f)=\Phi_{ij}(t=t_{\rm f}),
\end{aligned}
\end{equation}
where $t_{\rm f}=t_{\rm c}-(5/256)\mathcal M^{-5/3}_{\rm chirp}(\pi f)^{-8/3}$ and $t_{\rm c} \in [0,10]~\rm yr$  is the coalescence time~\cite{Maggiore:2007ulw}.

The term $t_{\rm f}$ represents the effect of the movement of the Earth during the time of the GW signal. If this effect is ignored, $t_{\rm f}$ can be treated as a constant. For binary coalescence, the duration of the signal $t_*$ is a strong function of the detector's low-frequency cutoff $f_{\rm lower}$~\cite{Maggiore:2007ulw},
\begin{equation}
	t_*=0.86~{\rm d} \left(\frac{1.21~M_{\odot}}{\mathcal{M}_{\rm chirp}}\right)^{5/3} \left(\frac{2~{\rm Hz}}{f_{\rm lower}}\right)^{8/3}.
\end{equation}
For 3G GW detectors, $f_{\rm lower}$ is extended to about 1 Hz. For the BNS with $m_1=m_2=1.4~M_{\odot}$, we have $t_*=5.44~\rm d$. Therefore, we cannot ignore the effect of the Earth's rotation. For this reason, we consider this effect in this paper (see, e.g., Ref.~\cite{Han:2023exn} for further discussion).

In each GW detection strategy, when the signal-to-noise ratio (SNR) exceeds the threshold of 12, we expect that the corresponding GW signal is detected. For low-mass systems, the combined SNR for the detection network of $N$ independent detectors is given by
\begin{equation}
\rho=(\tilde{\boldsymbol h}|\tilde{\boldsymbol h})^{1/2},
\end{equation}
where $\tilde{\boldsymbol h}$ is the frequency-domain GW waveform of $N$ independent detectors as mentioned in Eq.~(\ref{eq:waveform}).
The inner product is defined as
\begin{equation}
(\boldsymbol a|\boldsymbol b)=2\int_{f_{\rm lower}}^{f_{\rm upper}}\{\boldsymbol a(f)\boldsymbol b(f)^*+\boldsymbol b(f)\boldsymbol a(f)^*\}{\rm d}f,
\end{equation}
where $\boldsymbol a$ and $\boldsymbol b$ are column matrices of the same dimension, $*$ represents conjugate transpose, $f_{\rm lower}$ is the lower cutoff frequency ($f_{\rm lower}=1$ Hz for ET and $f_{\rm lower}=5$ Hz for CE), and $f_{\rm upper}=2/(6^{3/2}2\pi M_{\rm obs})$ is the upper cutoff frequency, with $M_{\rm obs}=(m_1+m_2)(1+z)$~\cite{Zhao:2010sz}.

\subsection{Detection of EM counterparts}

In this section, we will briefly review the operational details of the EM detections. We consider two types of EM counterparts to BNS mergers: GRBs and kilonovae. Following Refs.~\cite{Belgacem:2019tbw,Han:2023exn}, we first briefly introduce the method of GRB detections, which is expected to occur shortly after the BNS merger. Then we focus on the kilonova detections by ToO follow-up observations of GW triggers.

\subsubsection{GRB detection}

According to the observations of GW170817/GRB170817A, the jet profile model of short GRBs is given by~\cite{Howell:2018nhu}\footnote{In fact, the jet profile of short GRBs remains subject to significant uncertainties; see, e.g., Ref.~\cite{Hayes:2019hso} for more detailed discussions.}
\begin{equation}
L_{\rm iso}(\theta_{\rm v})=L_{\rm on}\exp\left(-\frac{\theta^2_{\rm v}}{2\theta^2_{\rm c}} \right),
\label{eq:jet}
\end{equation}
where $L_{\rm iso}(\theta_{\rm v})$ is the isotropically equivalent luminosity of short GRBs observed at different viewing angles $\theta_{\rm v}$, $L_{\rm on}$ is the on-axis isotropic luminosity defined by $L_{\rm on} \equiv L_{\rm iso}(0)$, and $\theta_{\rm c}$ is the characteristic angle of the core given by $\theta_{\rm c}=4.7^{\circ}$. Here we assume the directions of the jets are aligned with the binary orbital angular momentum, namely $\iota=\theta_{\rm v}$.

For the luminosity distribution of the short GRBs, we assume an empirical broken-power-law luminosity function~\cite{Wanderman:2009es}\footnote{The luminosity function of short GRBs is also uncertainties; see, e.g., Ref.~\cite{Salafia:2023sjx} for more detailed discussions.}
\begin{equation}
	\Phi(L)\propto
	\begin{cases}
		(L/L_*)^{\alpha_{\rm L}}, & L<L_*, \\
		(L/L_*)^{\beta_{\rm L}}, & L\ge L_*,
	\end{cases}
	\label{eq:LF}
\end{equation}
where $L$ is the peak luminosity of each burst, $L_{*}$ is the characteristic luminosity separating the low and high end of the luminosity function, and $\alpha_{\rm L}$ and $\beta_{\rm L}$ are the characteristic slopes describing these regimes. We set $\alpha_{\rm L}=-1.95$, $\beta_{\rm L}=-3$, and $L_{*}=2\times10^{52}~\rm erg~sec^{-1}$ following Ref.~\cite{Wanderman:2014eza}. In this paper, we refer to the on-axis isotropic luminosity $L_{\rm on}$ as the peak luminosity $L$ and assume a standard low-end cutoff in luminosity of $L_{\rm min} = 10^{49}~\rm erg~sec^{-1}$.

To determine the detection probability of a short GRB by sampling $\Phi(L){\rm d}L$, we use the standard flux-luminosity relation with two corrections, an energy normalization
\begin{equation}
C_{\rm det}=\frac{\int^{10000~\rm{keV}}_{1~\rm{keV}}EN(E){\rm d}E}{\int^{E_{\rm max}}_{E_{\rm min}}N(E){\rm d}E},
\end{equation}
and a $k$-correction
\begin{equation}
k(z)=\frac{\int^{E_{\rm max}}_{E_{\rm min}}N(E){\rm d}E}{\int^{E_{\rm max}(1+z)}_{E_{\rm min}(1+z)}N(E){\rm d}E},
\end{equation}
to convert the flux limit of the GRB detector $P_{\rm lim}$ to the isotropically equivalent luminosity $L_{\rm iso}$. Here, $[E_{\rm min},E_{\rm max}]$ is the detector's energy window~\cite{Howell:2018nhu,Wanderman:2014eza}. The observed photon flux is scaled by $C_{\rm det}$ to account for the missing fraction of the $\gamma$-ray energy seen in the detector band. The cosmological $k$-correction is due to the redshifted photon energy when traveling from source to detector. $N(E)$ is the observed GRB photon spectrum in units of $\rm{ph~s^{-1}~keV^{-1}~cm^{-2}}$.
For short GRB, the function $N(E)$ is simulated by the band function~\cite{Band:2002te}
\begin{widetext}
\begin{equation}
N(E)=\begin{cases}
N_0\left(\frac{E}{100~\rm keV}\right)^{\alpha_{\rm B}}\exp(-\frac{E}{E_0}),&E\leq E_{\rm b}, \\
 \\N_0\left(\frac{E_{\rm b}}{100~\rm keV}\right)^{\alpha_{\rm B}-\beta_{\rm B}}\exp(\beta_{\rm B}-\alpha_{\rm B}) \left(\frac{E}{100~\rm keV}\right)^{\beta_{\rm B}},&E>E_{\rm b},
\end{cases}
\end{equation}
\end{widetext}
where $E_{\rm b}=(\alpha_{\rm B}-\beta_{\rm B})E_0$ and $E_{\rm p}=(\alpha_{\rm B}+2)E_0$. Following Ref.~\cite{Wanderman:2014eza}, we adopt $\alpha_{\rm B}=-0.5$, $\beta_{\rm B}=-2.25$, and a peak energy $E_{\rm p}=800~\rm keV$ in the source frame. Then we can convert the flux limit $P_{\rm lim}$ to the luminosity $L_{\rm iso}$ by
\begin{equation}
L_{\rm iso}=4\pi d_{\rm L}^2(z)k(z)C_{\rm det}/(1+z)P_{\rm lim},
\end{equation}
according to the relation between flux and luminosity for GRBs~\cite{Meszaros:1995dj,Meszaros:2011zr}. Finally, using Eq.~(\ref{eq:LF}), we can select the detectable GRB from the BNS samples by sampling $\Phi(L){\rm d}L$.

\subsubsection{Kilonova detection}

Kilonovae are EM transients with durations ranging from a day to a week following BNS mergers. They are powered by the radioactive decay of r-process nuclei synthesized in the merger ejecta~\cite{Metzger:2010sy,Ascenzi:2018mbh,Rossi:2019fnm}. However, detecting kilonovae can be challenging due to their relatively low luminosity and rapidly evolving nature. To date, with the exception of AT2017gfo, which has relatively complete observations of kilonova properties, no kilonova candidate has been detected during the O3 observing run~\cite{Kasliwal:2020wmy,Becerra:2021lyq,deJaeger:2021tcq,Dichiara:2021vjy,Mohite:2021pfn}. Therefore, in this paper, we simply employ an AT2017gfo-like model as our standard kilonova model to calculate the kilonova detectability of GW-triggered ToO observations in view of the scarce data to date. Here, we adopt the \texttt{POSSIS} package to obtain the kilonova light curves~\cite{Bulla:2019muo}, with a total ejecta mass $0.04~M_{\odot}$ and a half-opening angle of the lanthanide-rich component $60^\circ$, which corresponds to the constraints by AT2017gfo~\cite{Cowperthwaite:2017dyu,Kasliwal:2017ngb,Villar:2017wcc,Murguia-Berthier:2017kkn,Perego:2017wtu,Tanaka:2017qxj,Zhu:2021zmy,Zhu:2021ram,Kang:2022nmz,Kasen:2017sxr}.\footnote{AT2017gfo provides a well-studied baseline, but kilonova emission remains model-dependent: variations in ejecta mass and composition, anisotropy with viewing angle, and uncertainties in nuclear heating can significantly affect light-curve amplitudes~\cite{Kasen:2017sxr,Barnes:2013wka}; consistent with this, observed kilonovae span a broad range in brightness, underscoring the intrinsic diversity of their emission~\cite{Gompertz:2017mbv,Prentice:2018qxn,Rossi:2019fnm,Rastinejad:2021nev}.}

Relatively bright jet afterglows could affect kilonova observations for on-axis or near-axis observers. Hence, we establish two criteria to define a detectable kilonova event~\cite{Zhu:2021ram,Kang:2022nmz}:
\begin{equation}
\begin{aligned}
	&({\rm i})~F_{\nu,\rm KN}(t_{\rm KN,p})+F_{\nu,\rm AG}(t_{\rm KN,p})>F_{\nu,\rm lim},~\rm and\\
	&({\rm ii})~F_{\nu,\rm KN}(t_{\rm KN,p})>5F_{\nu,\rm AG}(t_{\rm KN,p}),
\end{aligned}
\end{equation}
where $t_{\rm KN,p}$ is the peak time of the kilonova, $F_{\nu,\rm KN}$ and $F_{\nu,\rm AG}$ are the peak fluxes of kilonova and afterglow, and $F_{\nu,\rm lim}$ is the limiting flux for different survey projects. In this paper, we adopt \texttt{AFTERGLOWPY}, an open-source \texttt{PYTHON} package~\cite{Ryan:2019fhz}, to model the afterglow light curves and obtain the $F_{\nu,\rm AG}$, with parameters given in Ref.~\cite{Ryan:2019fhz}.

For the GW-triggered ToO observations of kilonovae, we apply a probabilistic statistical method to estimate the kilonova detection probability for BNS merger. The probability that a single kilonova event can be detected could be considered as the ratio of the survey area during the time duration $\Delta t$ when the brightness of the associated kilonova signal exceeds the limiting flux $F_{\nu,\rm lim}$ to the area of the sky localization of the relevant GW event $\Omega_{\rm GW}$, which can be obtained in the next subsection. The maximum probability for a source to be detected is $\Omega_{\rm FoV}\Delta t/[\Omega_{\rm GW}(t_{\rm exp}+t_{\rm oth})]$, where $\Omega_{\rm FoV}$ is the FoV for different survey projects, $t_{\rm exp}$ is the exposure time, and $t_{\rm oth}$ is the rest of time spent for each visit, which depends on the technical performance of the special survey project and the different search strategies. Since $t_{\rm oth}$ is uncertain, we treat it as a constant for each survey project in our calculations, i.e., $t_{\rm oth}=15~\rm s$.

Nevertheless, the event must fall within the sky coverage of the survey projects to have a chance of being discovered. Therefore, we set an upper limit on the probability of detecting a source, which is given by $\Omega_{\rm cov}/\Omega_{\rm sph}$. Here, $\Omega_{\rm cov}$ is the detectable sky coverage for a survey project, and $\Omega_{\rm sph}=41252.96~\rm deg^2$ is the area of the celestial sphere. Therefore, the kilonova detection probability can be written as
\begin{equation}
	P_{\rm KN}=\frac{\Omega_{\rm cov}}{\Omega_{\rm sph}}\times\frac{\Omega_{\rm FoV}{\rm min}(t_{\rm cad},\Delta t)}{{\rm max}(\Omega_{\rm FoV},\Omega_{\rm GW})(t_{\rm exp}+t_{\rm oth})},
\end{equation}
where $t_{\rm cad}$ is the cadence time for each event related to $t_{\rm exp}$ and $\Omega_{\rm GW}$.

\subsection{Fisher information matrix and error analysis}

We use the Fisher information matrix (FIM) to estimate the instrumental error of
GWs. The FIM of a GW detector network is given by
\begin{equation}
F_{ij}=\left(\frac{\partial \tilde{\boldsymbol{h}}}{\partial \theta_i}\Bigg |\frac{\partial \tilde{\boldsymbol{h}}}{\partial \theta_j}\right),
\end{equation}
where $\boldsymbol\theta$ denotes nine GW source parameters ($d_{\rm L}$, $\mathcal{M}_{\rm chirp}$, $\eta$, $\theta$, $\phi$, $\iota$, $t_{\rm c}$, $\psi_{\rm c}$, $\psi$). The covariance matrix is equal to the inverse of the Fisher matrix, i.e., ${\rm Cov}_{ij}=(F^{-1})_{ij}$. Thus, the instrumental error of GW parameter $\theta _i$ is $\Delta\theta_i=\sqrt{{\rm Cov}_{ii}}$.

The sky localization error can be given as
\begin{equation}
	\Omega_{\rm GW}=2\pi|\sin\theta|\sqrt{(\Delta \theta)^2(\Delta\phi)^2-(\Delta\theta\Delta\phi)^2}.
\end{equation}

The total error of the luminosity distance $d_{\rm L}$  includes the instrumental error $\sigma_{d_{\mathrm{L}}}^{\mathrm{inst}}$ estimated by the FIM, the weak-lensing error $\sigma_{d_{\mathrm{L}}}^{\mathrm{lens}}$, and the peculiar velocity error $\sigma_{d_{\mathrm{L}}}^{\mathrm{pv}}$~\cite{Jin:2021pcv,Wu:2022dgy,Jin:2022tdf,Jin:2023zhi,Jin:2023tou}. It can be written as
\begin{equation}
	\left(\sigma_{d_{\mathrm{L}}}\right)^{2}=\left(\sigma_{d_{\mathrm{L}}}^{\mathrm{inst}}\right)^{2}+\left(\sigma_{d_{\mathrm{L}}}^{\text {lens }}\right)^{2}+\left(\sigma_{d_{\mathrm{L}}}^{\text {pv }}\right)^{2}.\label{eq:total}
\end{equation}
The error caused by weak lensing is given in Refs.~\cite{Hirata:2010ba,Tamanini:2016zlh,Speri:2020hwc},
\begin{equation}
\sigma_{d_{\rm L}}^{\rm lens}(z)=d_{\rm L}(z) \times 0.066\left [\frac{1-(1+z)^{-0.25}}{0.25}\right ]^{1.8}.\label{lens}
\end{equation}
The error caused by the peculiar velocity is obtained from Ref.~\cite{Kocsis:2005vv}
\begin{equation}
	\sigma_{d_{\rm L}}^{\rm pv}(z)=d_{\rm L}(z)\times \left [1+ \frac{c(1+z)^2}{H(z)d_{\rm L}(z)}\right ]\frac{\sqrt{\langle v^2\rangle}}{c},
\end{equation}
where $H(z)$ is the Hubble parameter, $c$ is the speed of light in vacuum, and $\sqrt{\langle v^2\rangle}$ is the peculiar velocity. To be consistent with the average value of the galaxy catalogs, we roughly set $\sqrt{\langle v^2\rangle}=500~{\rm km~s^{-1}}$~\cite{He:2019dhl}.

In this paper, we treat the GW events from the GW-GRB, GW-kilonova, and their combined joint GW-EM detections as standard sirens for the cosmological analysis. We employ the Markov chain Monte Carlo (MCMC) analysis using the \texttt{PYTHON} package \texttt{EMCEE}~\cite{Lewis:2002ah,Foreman-Mackey:2012any}, and maximize the likelihood $\mathcal{L}\propto(-\chi^2/2)$ to infer the posterior probability distributions of cosmological parameters $\vec{\Omega}$. The $\chi^2$ function is expressed as
\begin{equation}
	\chi^2=\sum\limits_{i=1}^N\left[\frac{{d}_{\rm L}^i-d_{\rm L}(z_i,\vec{\Omega})}{{\sigma}_{d_{\rm L}}^i}\right]^2,
\end{equation}
where ${z}_i$, ${d}_{\rm L}^i$, and ${\sigma}_{d_{\rm L}}^i$ denote the $i$-th GW event's redshift, luminosity distance, and the total error of the luminosity distance, respectively.

\section{Multi-messenger Observations of GWs and EM counterparts}\label{sec3}

 In this section, we will discuss the BNS mergers' detection rates and distributions of GW observations, short GRB observations, and GW-triggered ToO observations of kilonovae (hereafter abbreviated as GW-kilonova observations).

 For the GW observations, we consider two different cases: the single ET and the ET-CE-CE network (one ET detector and two CE-like detectors, one in the USA with a 40 km arm length and another one in Australia with a 20 km arm length, hereafter abbreviated as ET2CE). We adopt the sensitivity curve of ET from Ref.~\cite{ETcurve-web} and the sensitivity curves of CE from Ref.~\cite{CEcurve-web}, as illustrated in Fig.~1 of Ref.~\cite{Han:2023exn}. For the GW detectors, in view of the high uncertainty of the duty cycle, we only calculate the best case where each detector has a duty cycle of 100\%~\cite{Zhu:2021ram}. The specific parameters characterizing the GW detector geometry (latitude $\varphi$, longitude $\lambda$, opening angle $\zeta$, and arm bisector angle $\gamma$) are listed in Table~{\uppercase\expandafter{\romannumeral 1}} of Ref.~\cite{Han:2023exn}.

For the GRB observations, we adopt GECAM with a sensitivity of $2\times 10^{-8} ~\rm erg~s^{-1}~cm^{-2}$ in the $6~\rm keV$--$5~\rm MeV$ band~\cite{Zhang:2018hzq}, considering that GECAM can achieve nearly all-sky coverage for detecting GRB events.

For the GW-kilonova observations, we consider three proposed survey projects using the most common filters ($g$, $r$, $i$): WFST, LSST, and CSST. Note that, without loss of generality, we choose $t_{\rm exp}=300~\rm s$ throughout this paper. The technical parameters of these survey projects are provided in Table~\ref{tab1}.

\begin{table*}
	\renewcommand\arraystretch{1.5}
	\centering
	\caption{ A summary of technical parameters for each survey project, based on Refs.~\cite{Zhu:2021ram,Zhu:2020ffa,Kang:2022nmz,2018AcASn..59...22S,LSSTScience:2009jmu,Gong:2019yxt}, assuming the exposure time $t_{\rm exp}=300~\rm s$ for the search limiting magnitude $m_{\nu,\rm lim}$.}
	\label{tab1}
	\setlength{\tabcolsep}{8.35mm}
	{\begin{tabular}{c ccc c c}
			\hline \hline
			\multirow{2}{*}{Telescope}        & \multicolumn{3}{c}{$m_{\nu,\rm lim}$}
			& \multirow{2}{*}{FoV/$\rm{deg}^2$}  & \multirow{2}{*}{Sky Coverage/$\rm{deg}^2$}   \\
			& $\it{g}$ & $\it{r}$ & $\it{i}$\\
			 \hline
			WFST       & 24.18        & 23.95        & 23.33        & 6.55 & 20,000 \\
			LSST       & 26.15        & 25.70        & 25.79        & 9.6     &  20,000\\
			CSST       & 26.30         & 26.00         & 25.90         & 1.1     &17,500    \\
			\hline \hline
	\end{tabular}}
\end{table*}

\begin{figure}[htbp]
	\includegraphics[width=0.9\linewidth,angle=0]{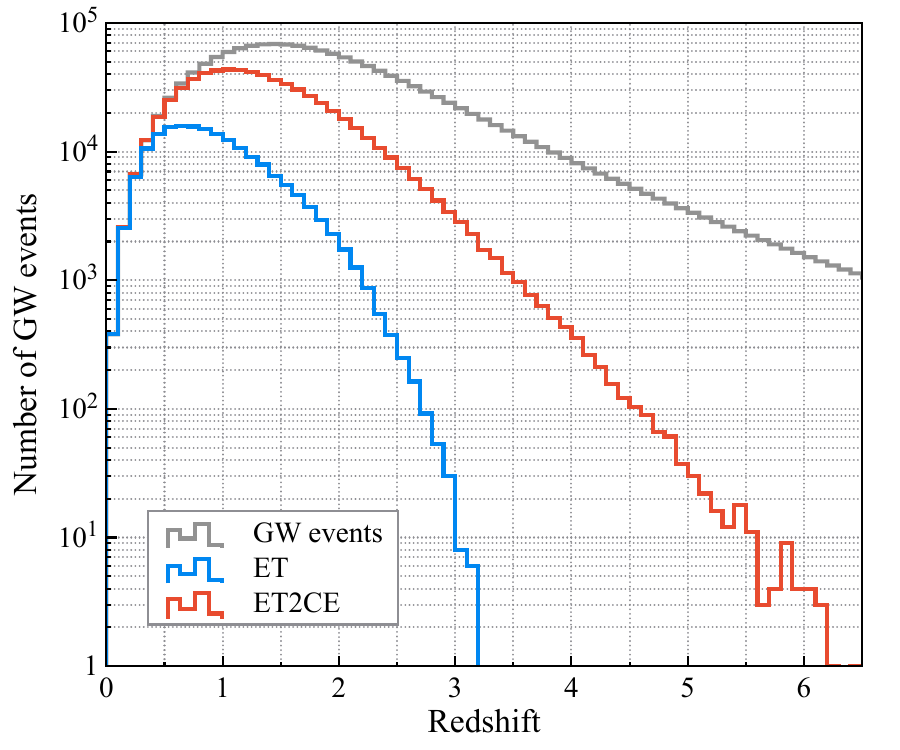}
	\caption{\label{fig1} Redshift distributions of all GW events and those detected by ET and ET2CE in a 10-year observation.}
\end{figure}

\begin{figure}[htbp]
	\includegraphics[width=0.9\linewidth,angle=0]{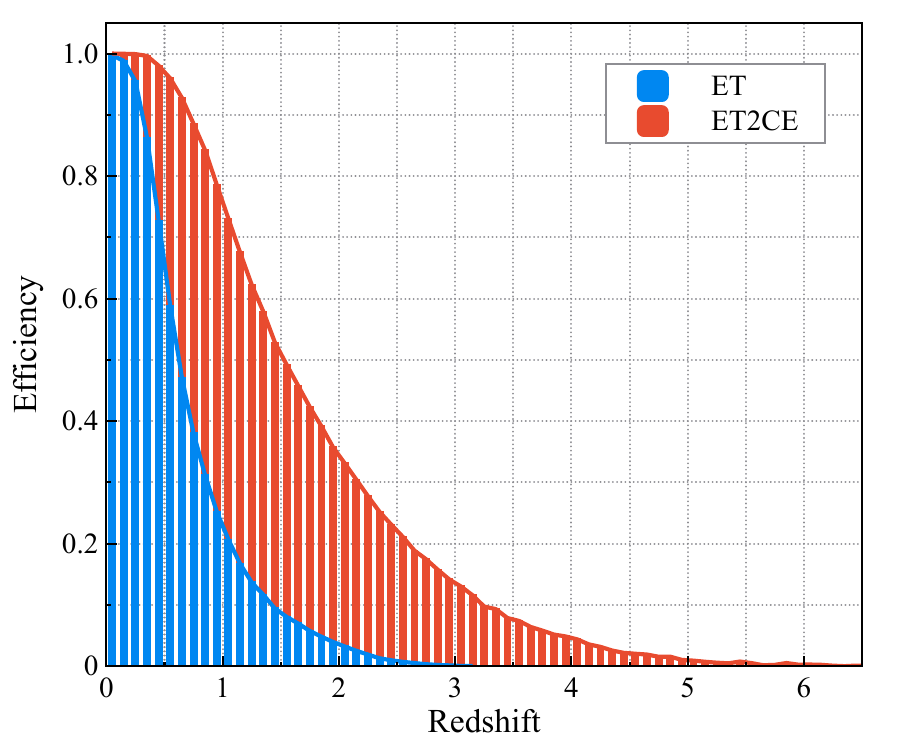}
	\caption{\label{fig2} Detection efficiencies of ET and ET2CE.}
\end{figure}

\begin{figure}[htbp]
	\includegraphics[width=0.9\linewidth,angle=0]{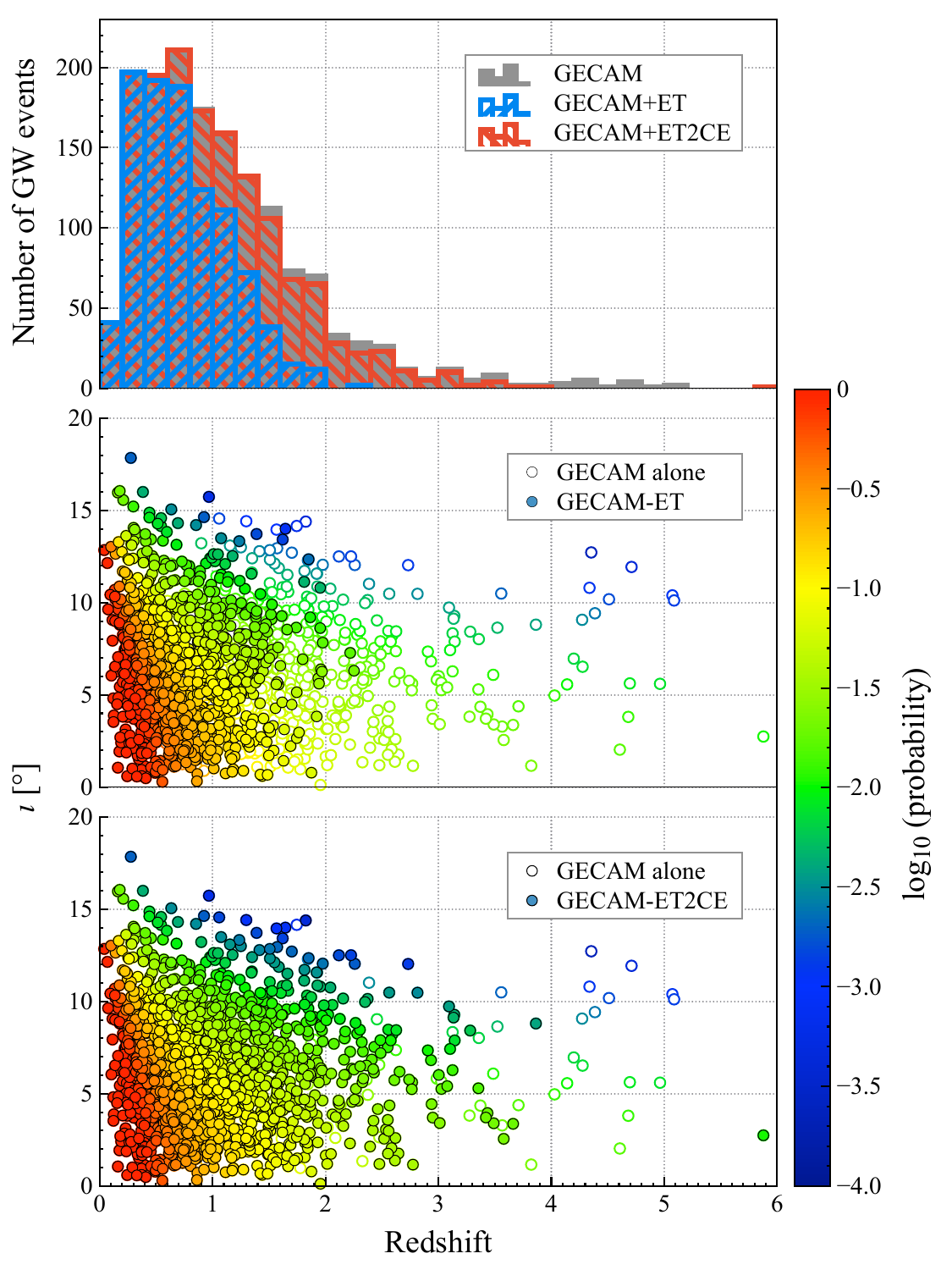}
	\caption{\label{fig3} Redshift distributions of detectable short GRBs and GW-GRB coincidences in a 10-year observation. {Top panel}: Redshift distributions of BNS mergers triggered by GECAM and GECAM in synergy with ET and ET2CE. {Lower two panels}: Distributions of inclination angles $\iota$ and redshifts of BNS mergers that can be triggered by GECAM alone and GECAM in synergy with ET and ET2CE. The color bar represents the logarithm of detection probability for GECAM.}
\end{figure}

We first calculate the redshift distributions of all GW events, as well as those detected by the ET and the ET2CE network, assuming a 10-year observation as illustrated in Fig.~\ref{fig1}, along with their detection efficiencies shown in Fig.~\ref{fig2}. The results indicate that the GW detector network can detect substantially more events than a single observatory. For the single ET, the horizon redshift is up to $\sim 3.2$, with $50\%$ detection efficiency at $\sim 0.6$. For the ET2CE network, the horizon redshift extends to $\sim 6.2$, with $50\%$ detection efficiency at $\sim 1.5$.

Then, we select the GW events that can be triggered by both GECAM telescope and GW detectors. In Fig.~\ref{fig3}, we show the redshift distributions of detectable short GRBs and GW-GRB coincidences for a 10-year observation. The top panel of Fig.~\ref{fig3} shows the redshift distributions of BNS mergers triggered by GECAM, as well as GECAM in synergy with ET and ET2CE. For the ET alone, $65.4\%$ of all the detectable GRBs are expected to have detectable GW counterparts. For the ET2CE network, the vast majority of detectable short GRBs have detectable GW counterparts, with joint detection efficiency reaching up to 96.0\%.

The lower two panels of Fig.~\ref{fig3} show the distributions of the inclination angles and the redshifts of BNS mergers, which can be triggered by GECAM alone and by GECAM in synergy with ET and ET2CE. The color bar represents the logarithm of the detection probability for GECAM. Due to the constraints imposed by the Gaussian jet profile, the GW events jointly detectable by both GRB and GW detectors are limited to inclination angles $\iota < 20^{\circ}$. As the redshift and inclination angle increase, the detection probability of GRBs decreases significantly. Under the intrinsic limitations of the structured Gaussian jet scenario, although the ET2CE network detects several times more sources than the single ET (and at much higher redshifts; see Fig.~\ref{fig1}), the corresponding number of joint GW-GRB detections does not increase proportionally. This is due to intrinsic limitations in GRB detections.

\begin{table*}[!htb]
	\caption{ Numbers of BNS events detected by ET, ET2CE, and GRB events triggered by GECAM in a 10-year observation and joint GW-GRB events triggered by GECAM in synergy with ET and ET2CE.}
	\label{tab2}
	\setlength\tabcolsep{5.88mm}
	\renewcommand{\arraystretch}{2}
	\begin{tabular}{ccccc}
		\hline \hline
		BNS samples&Detection strategy&GW detections	&GRB detections	&GW-GRB detections\\
		\hline
		\multirow{2}{*}{1,512,710}&      ET       &   180,029    &     \multirow{2}{*}{1516}    &  992    \\
	    \multirow{2}{*}{}              &    ET2CE   &   661,704 &     \multirow{2}{*}{}         &      1455  \\
		\hline\hline
	\end{tabular}
\end{table*}

In Table~\ref{tab2}, we show the results of our simulations for the 3G era in terms of the number of GW and GRB detections from BNS mergers, as well as the number of joint GW-GRB detections assuming a 10-year observation. For the single ET detector, our estimate of GW detections is $\sim 1.80\times 10^5$ in 10 years. This number is lower than the estimate in our previous work~\cite{Han:2023exn}, primarily because the local comoving merger rate and the delay time distribution between the formation and merger of BNS systems assumed in this paper differ from those in Ref.~\cite{Han:2023exn}. For the GRB detections, we estimate that 1516 short GRBs could be triggered by GECAM. Furthermore, with the single ET detector, we expect 992 coincident GW-GRB events with a redshift distribution below $\sim 2.4$ in a 10-year observation, consistent with the order of magnitude reported in previous works~\cite{Hou:2022rvk,Han:2023exn,Chen:2020zoq,Yu:2021nvx}. With the ET2CE network, the number of detectable GW-GRB events increases to 1455, extending the redshift range to $\sim 4$.

\begin{figure*}[htbp]
	\includegraphics[width=0.45\linewidth,angle=0]{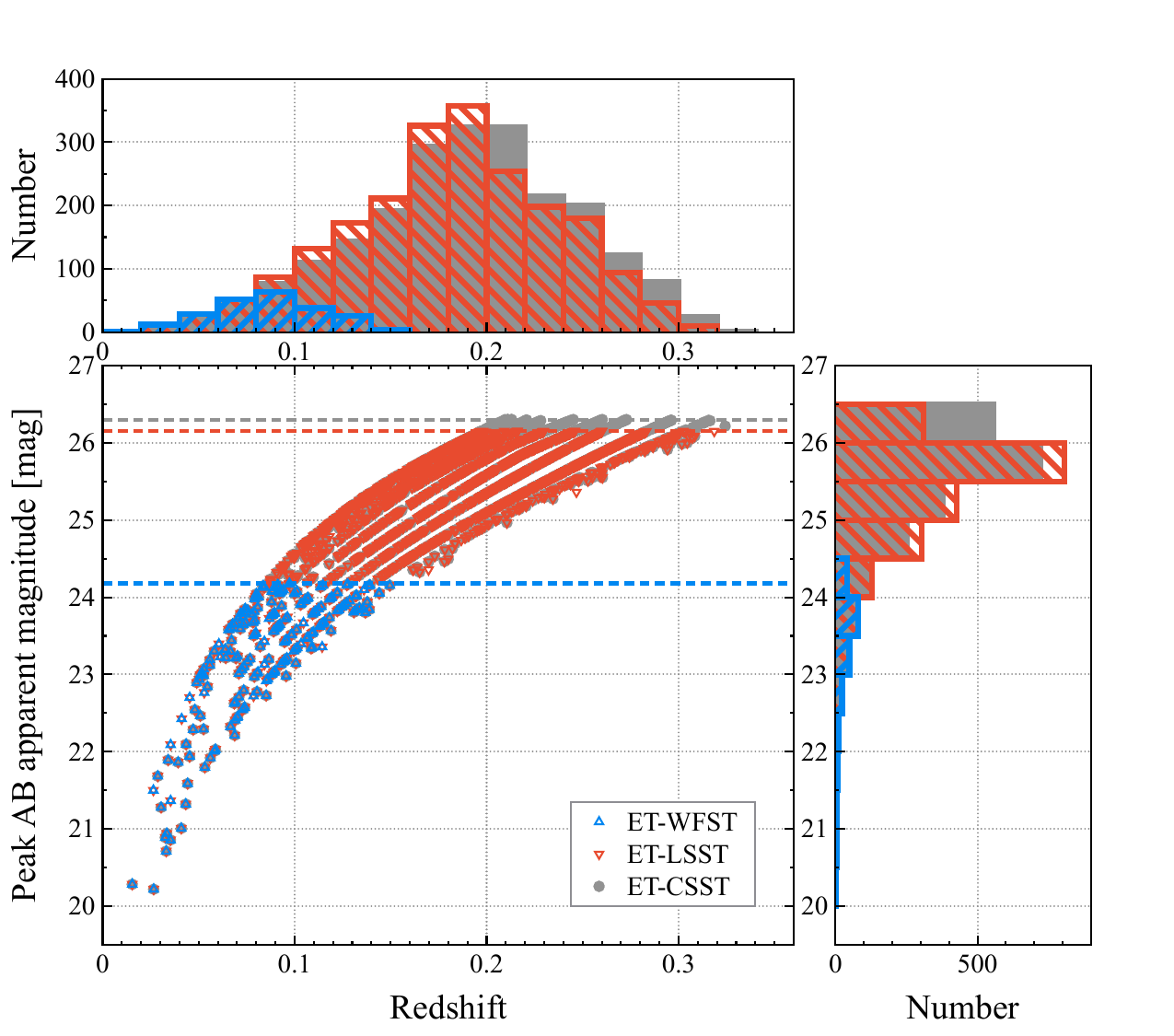}
	\includegraphics[width=0.45\linewidth,angle=0]{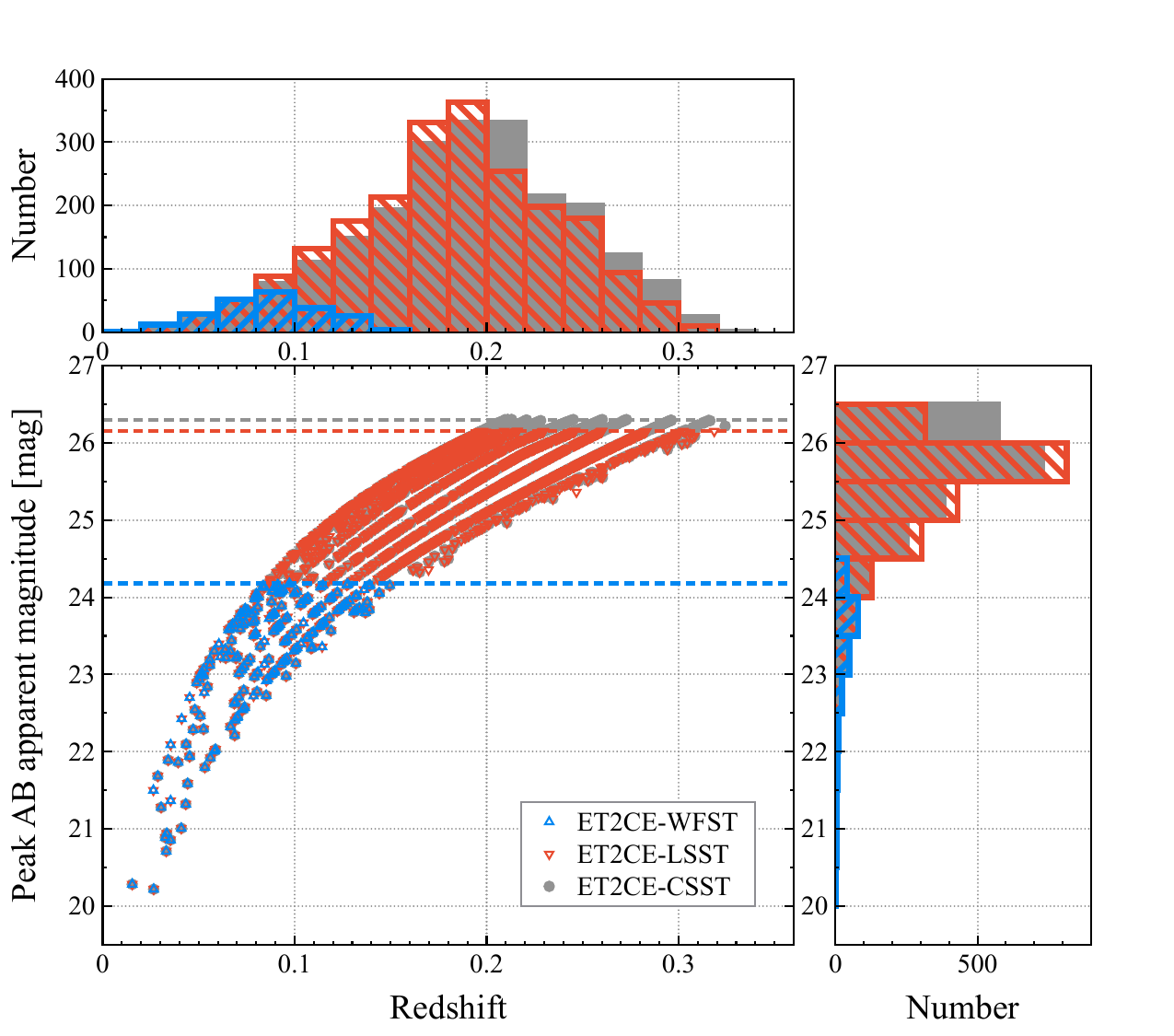}
	\caption{\label{fig4}  Redshift distributions with respect to the peak AB apparent magnitude for GW-kilonova detections in the $g$ band under different survey projects assuming a 10-year observation. The left panel shows the results of ET and the right panel shows those of ET2CE. In each panel, circles with different colors represent different GW-kilonova detections, and dashed horizontal lines with different colors denote the search limiting magnitude  $m_{\nu,\rm lim}$ of the corresponding survey projects (see Table~\ref{tab1}).}
\end{figure*}

\begin{figure*}[htbp]
	\includegraphics[width=0.45\linewidth,angle=0]{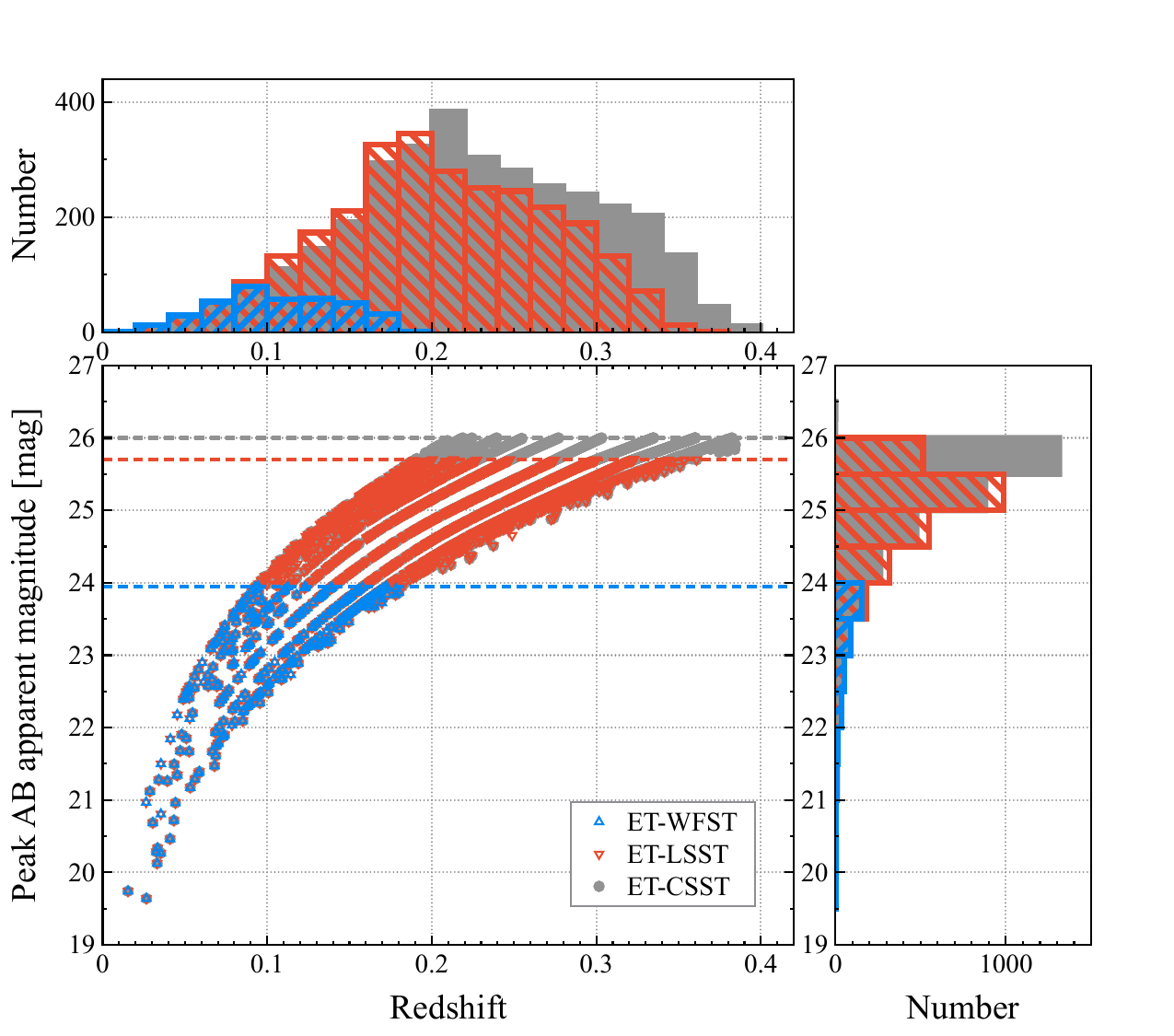}
	\includegraphics[width=0.45\linewidth,angle=0]{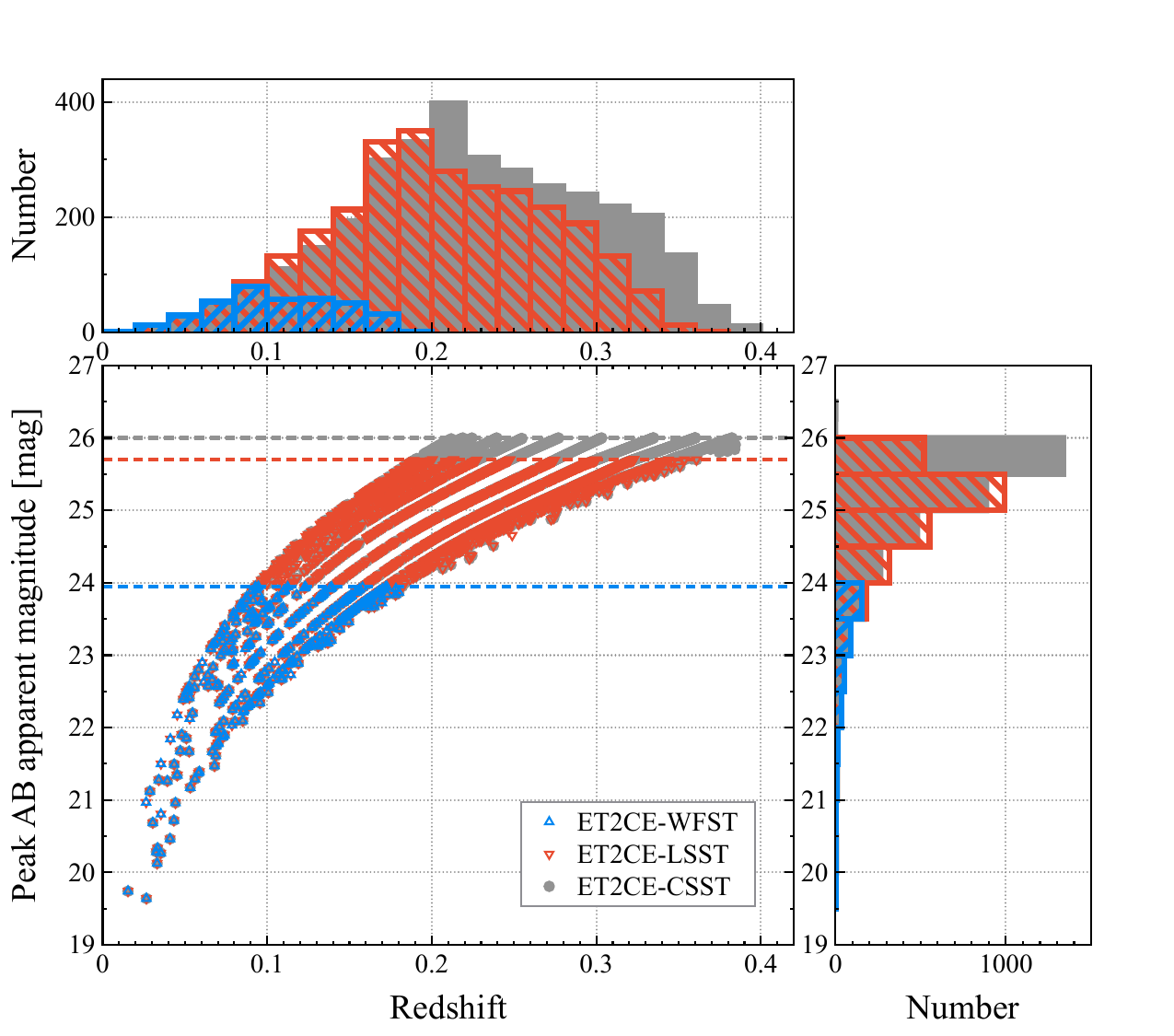}
	\caption{\label{fig5} Same as Fig.~\ref{fig4}, but assuming GW-kilonova detections in the $r$ band.}
\end{figure*}

\begin{figure*}[htbp]
	\includegraphics[width=0.45\linewidth,angle=0]{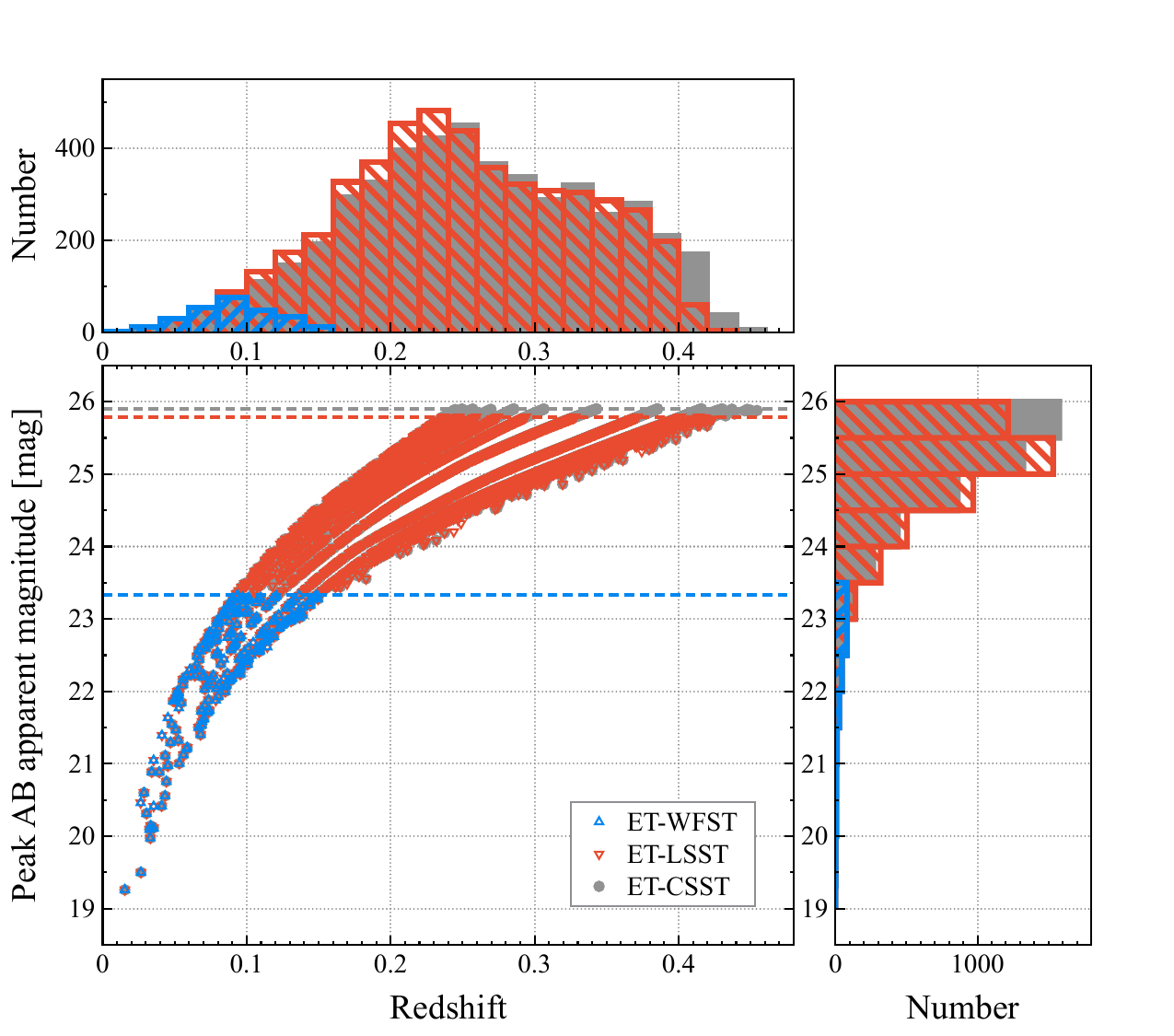}
	\includegraphics[width=0.45\linewidth,angle=0]{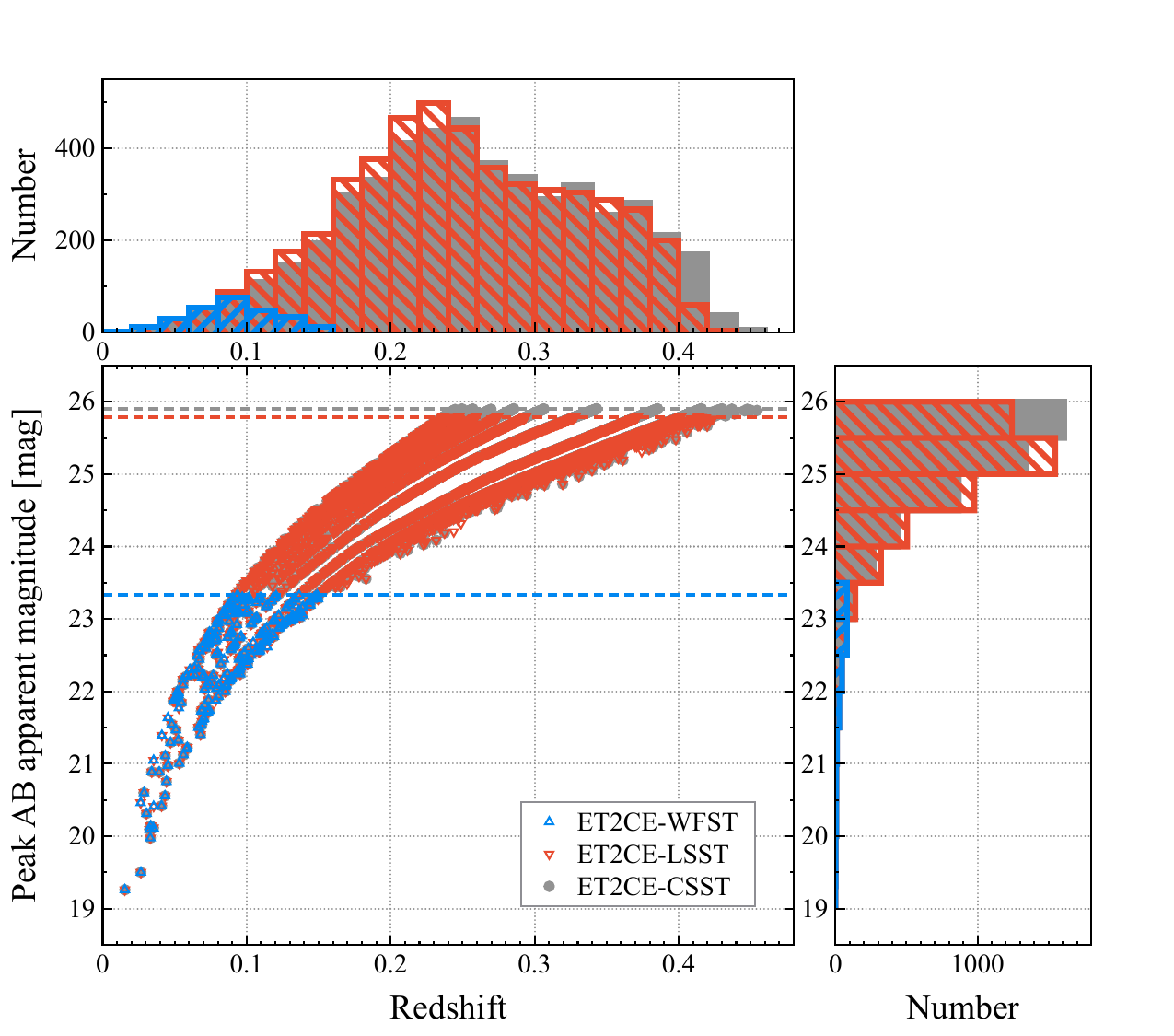}
	\caption{\label{fig6} Same as Fig.~\ref{fig4}, but assuming GW-kilonova detections in the $i$ band.}
\end{figure*}

\begin{table*}
	\renewcommand\arraystretch{1.5}
	\centering
	\caption{ Numbers of GW-kilonova detections in the three bands under different survey projects in a 10-year observation.}
	\label{tab3}
	\setlength{\tabcolsep}{3.63mm}
	{\begin{tabular}{c|ccc|ccc|ccc}
			\hline \hline
			\multirow{2}{*}{Detection strategy}  & \multicolumn{3}{c|}{$\it{g}$} & \multicolumn{3}{c|}{$\it{r}$}  & \multicolumn{3}{c}{$\it{i}$}\\
			 & WFST   & LSST & CSST & WFST & LSST & CSST & WFST & LSST & CSST \\
			\hline
			ET              & 225  & 2162  & 2183  & 372   &2766   &3288  &264   &4863  &4752 \\
			ET2CE       & 225  & 2178  & 2209  & 372   &2782   &3321 &264  &4917   &4816\\
			\hline \hline
	\end{tabular}}
\end{table*}

For the GW-kilonova detections, we present the redshift distributions with respect to the peak AB apparent magnitude, given specific combinations of GW detectors and optical survey projects in different filters ($g$, $r$, $i$) assuming a 10-year observation in Figs.~\ref{fig4}--\ref{fig6}. More detailed results are listed in Table~\ref{tab3}. We can see that, compared to the GW-GRB detections in Fig.~\ref{fig3}, the GW-kilonova detections have lower redshifts. Meanwhile, considering different $m_{\nu,\rm lim}$ for optical survey projects in each of the bands, it is observed that the horizon redshifts of GW-kilonova detections increase with the improved sensitivity of the survey projects from WFST to CSST in the three bands. This can be understood as more GW-kilonova events will be detected at higher redshifts from WFST to CSST. In addition, LSST in the $i$ band gives the highest number of detections among the three bands across different survey projects, with $\sim 4900$ events detected at redshifts below $\sim 0.4$, more than three times that of GW-GRB detections under ET2CE. It should be noted that, although the $m_{\nu,\rm lim}$ of CSST is higher than that of LSST in all three bands, the numbers of GW-kilonova detections of CSST in the $i$ band are lower than those of LSST. The main reason is that, although CSST has higher sensitivity than LSST, its performance is constrained by its more limited sky coverage. Finally, we find that ET and ET2CE produce almost identical GW-kilonova detections when using the same survey project in the same band. This is primarily because, at redshifts below $\sim 0.4$, ET and ET2CE exhibit comparable detection efficiencies, as illustrated in Fig.~\ref{fig2}, with minimal effect on the following kilonova detections.

We emphasize that the detection numbers reported in Tables~\ref{tab2} and~\ref{tab3} are derived from a deterministic threshold-based selection applied to different types of detectors and represent a single realization of our simulation pipeline. While no formal statistical errors are assigned, these numbers should be regarded as subject to significant sources of error arising from modeling assumptions, including, but not limited to, the intrinsic BNS merger rate, the GRB emission model (e.g., jet profile and luminosity function), and the kilonova emission model (e.g., ejecta mass and composition, anisotropy with viewing angle, and nuclear heating rate). A more comprehensive quantification of these uncertainties is beyond the scope of this work, but we stress that such an investigation will be an important direction for future work.

\section{Constraint results on cosmological parameters}\label{sec4}

In this section, we shall report the constraint results on cosmological parameters. We consider the $\Lambda$CDM, $w$CDM, and $w_0w_a$CDM models to perform cosmological analysis using the MCMC method. We constrain these cosmological models using GW multi-messenger detections with different EM detection scenarios, including GW-GRB, GW-kilonova, and their combined GW-EM detections. Meanwhile, we also give the constraint results for the CMB+BAO+SN dataset (hereafter abbreviated as CBS) and CBS combined with those GW multi-messenger detections to show the capability of GW multi-messenger observations in breaking the cosmological parameter degeneracies. For the CMB data, we employ the ``Planck distance priors'' from the Planck 2018 observation~\cite{Chen:2018dbv}. For the BAO data, we adopt the measurements from 6dFGS ($z_{\rm eff}=0.106$)~\cite{Beutler:2011hx}, SDSS-MGS ($z_{\rm eff}=0.15$)~\cite{Ross:2014qpa}, and BOSS DR12 ($z_{\rm eff}=0.38$, 0.51, and 0.61)~\cite{BOSS:2016wmc}. For the SN data, we employ the ``Pantheon+'' compilation~\cite{Brout:2022vxf}. The 1$\sigma$ and 2$\sigma$ posterior distribution contours for the cosmological parameters of interest are shown in Figs.~\ref{fig7},~\ref{fig9}--\ref{fig14} and the 1$\sigma$ errors for the marginalized parameter constraint results are shown in Tables~\ref{tab4},~\ref{tab5}, and~\ref{tab7}--\ref{tab10}. We use $\sigma(\xi)$ and $\varepsilon(\xi)$ to represent the absolute and relative errors of parameter $\xi$, with $\varepsilon(\xi)$ defined as $\varepsilon(\xi)=\sigma(\xi)/\xi$.

\subsection{Constraint results from GW multi-messenger observations with different EM detection scenarios}

\begin{figure*}[htbp]
	\includegraphics[width=0.4\linewidth,angle=0]{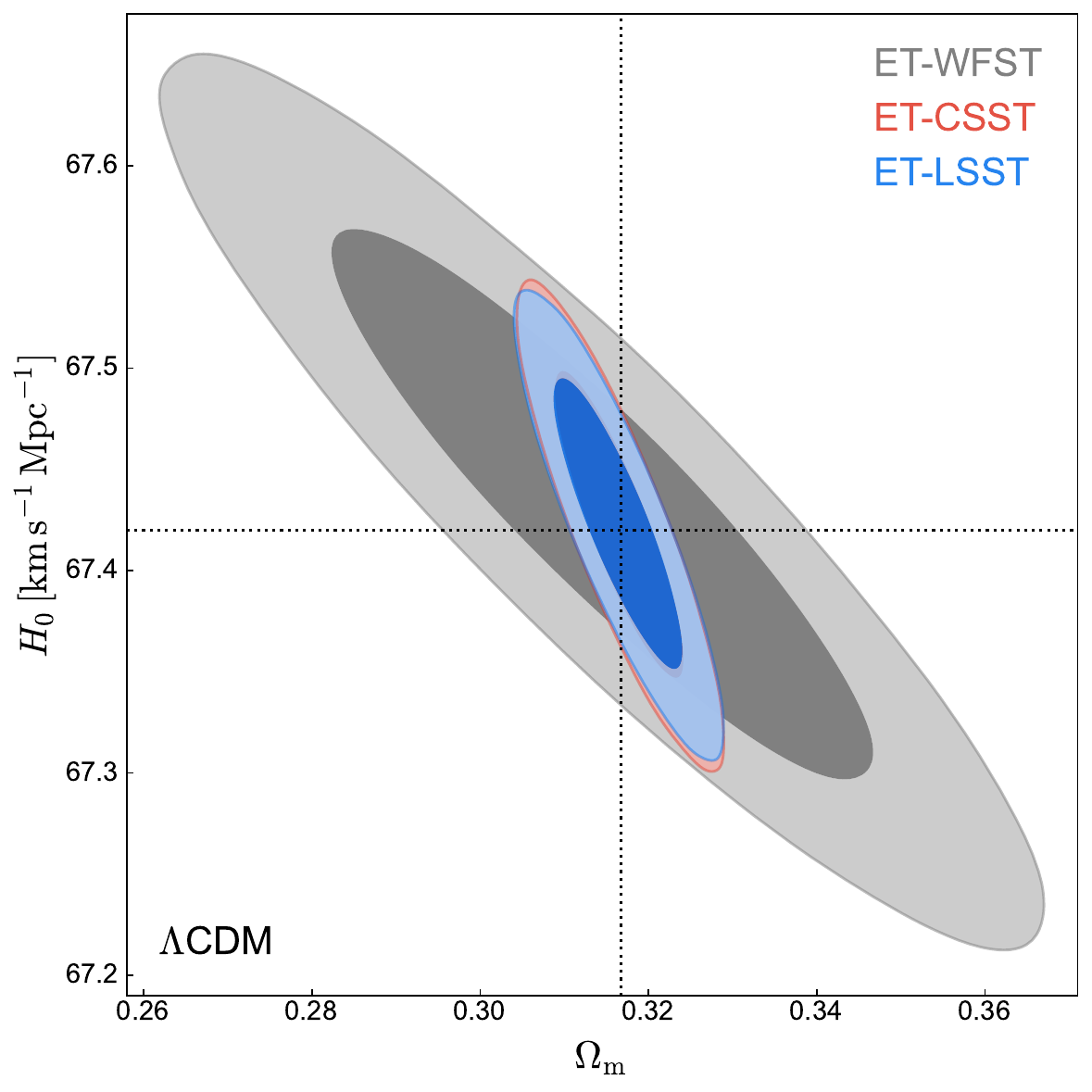}
	\includegraphics[width=0.4\linewidth,angle=0]{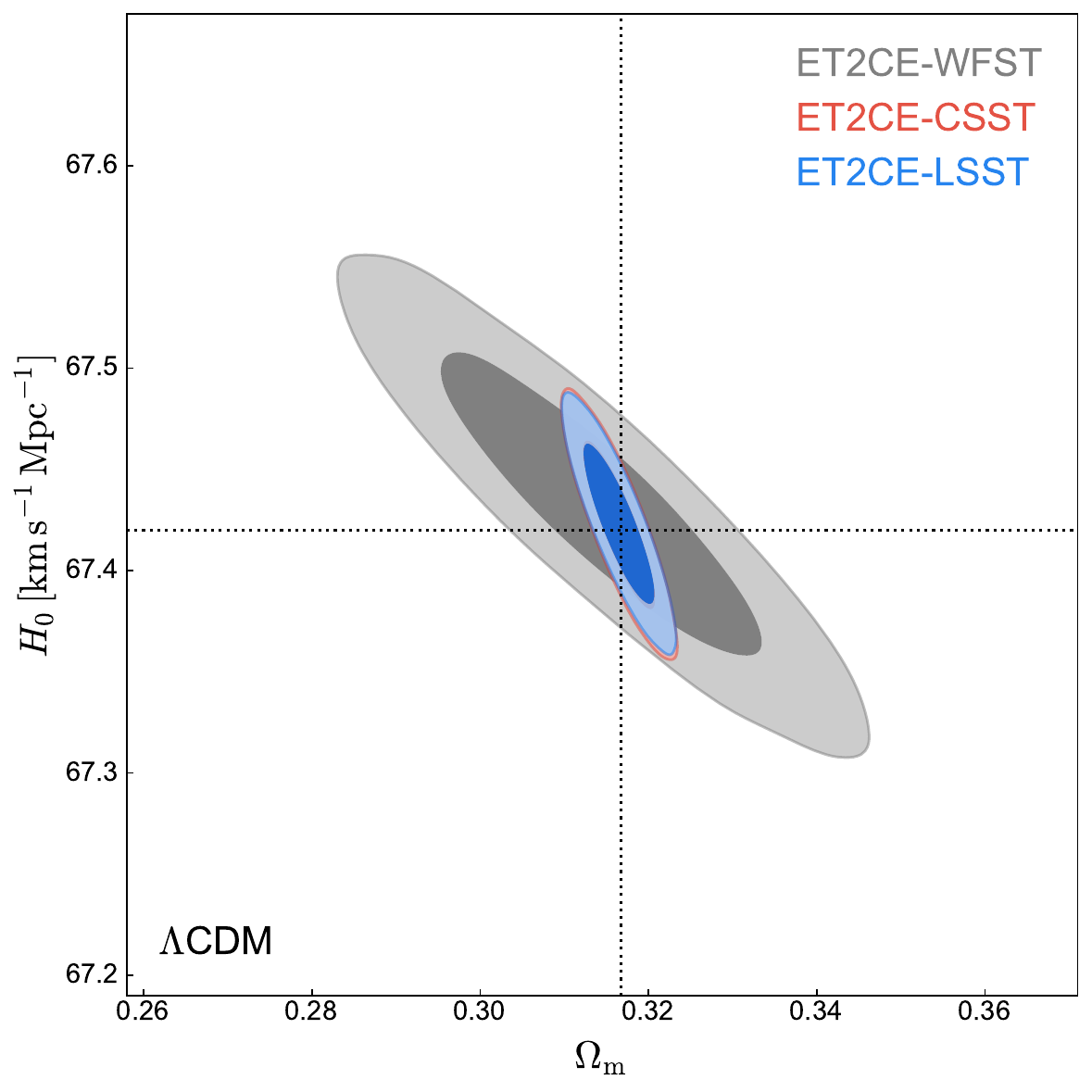}
	\caption{\label{fig7} Two-dimensional marginalized contours ($68.3\%$ and $95.4\%$ confidence level) in the $\Omega_{\rm m}$--$H_0$ plane for the $\Lambda$CDM model using GW-kilonova detections in the $i$ band under different survey projects. The left panel shows the constraint results from ET and the right panel shows those from ET2CE. Here, the dotted lines indicate the fiducial values of cosmological parameters preset in the simulation.}
\end{figure*}

\begin{table*}
	\renewcommand\arraystretch{1.5}
	\centering
	\caption{ Absolute (1$\sigma$) and relative errors of cosmological parameters in the $\Lambda$CDM model using GW-kilonova detections in the three bands under different survey projects. Here, $H_0$ is in units of $\rm km\ s^{-1}\ Mpc^{-1}$.}
	\label{tab4}
	\setlength{\tabcolsep}{2.85mm}
	{\begin{tabular}{c|c|ccc|ccc|ccc}
			\hline \hline
			\multirow{2}{*}{\shortstack{Detection \\ strategy}} &\multirow{2}{*}{Error}&\multicolumn{3}{c|}{$\it{g}$} & \multicolumn{3}{c|}{$\it{r}$}  & \multicolumn{3}{c}{$\it{i}$}\\
			 &&WFST   & LSST & CSST & WFST & LSST & CSST & WFST & LSST & CSST \\
			\hline
			\multirow{4}{*}{ET}
			                  &$\sigma(\Omega_{\rm m})$&$0.0230$  &$0.0080$  &$0.0081$   &$0.0180$  &$0.0068$  &$0.0063$  &$0.0220$     &$0.0051$   &$0.0051$\\
			                  &$\sigma(H_0)$                     &$0.093$    &$0.058$    &$0.060$     &$0.083$    &$0.054$    &$0.054$     &$0.091$     &$0.047$    &$0.049$\\
							  &$\ve(\Omega_{\rm m})$      &$7.32\%$  &$2.53\%$  &$2.56\%$   &$5.71\%$   &$2.15\%$   &$1.99\%$   &$7.01\%$ &$1.61\%$   &$1.61\%$\\
							  &$\ve(H_0)$                           &$0.138\%$&$0.086\%$&$0.089\%$&$0.123\%$&$0.080\%$&$0.080\%$&$0.135\%$ &$0.070\%$&$0.073\%$\\
			\hline
			\multirow{4}{*}{ET2CE}
			                  &$\sigma(\Omega_{\rm m})$&$0.0130$   &$0.0043$  &$0.0043$  &$0.0110$  &$0.0037$   &$0.0034$  &$0.0125$   &$0.0028$  &$0.0028$\\
							  &$\sigma(H_0)$                     &$0.050$     &$0.031$     &$0.033$    &$0.047$    &$0.030$    &$0.030$     &$0.050$     &$0.027$    &$0.028$\\
							  &$\ve(\Omega_{\rm m})$      &$4.14\%$   &$1.36\%$   &$1.36\%$   &$3.49\%$  &$1.17\%$    &$1.07\%$   &$3.99\%$    &$0.88\%$  &$0.88\%$\\
							  &$\ve(H_0)$                           &$0.074\%$&$0.046\%$&$0.049\%$&$0.070\%$&$0.044\%$&$0.044\%$&$0.074\%$&$0.040\%$&$0.042\%$\\
			\hline \hline
	\end{tabular}}
\end{table*}

\begin{table*}
	\renewcommand\arraystretch{1.5}
	\centering
	\caption{ Same as Table~\ref{tab4}, but in the $w$CDM model.}
	\label{tab5}
	\setlength{\tabcolsep}{2.85mm}
	{\begin{tabular}{c|c|ccc|ccc|ccc}
			\hline \hline
			\multirow{2}{*}{\shortstack{Detection \\ strategy}} &\multirow{2}{*}{Error}&\multicolumn{3}{c|}{$\it{g}$} & \multicolumn{3}{c|}{$\it{r}$}  & \multicolumn{3}{c}{$\it{i}$}\\
			&&WFST   & LSST & CSST & WFST & LSST & CSST & WFST & LSST & CSST \\
			\hline
			\multirow{6}{*}{ET}
			&$\sigma(\Omega_{\rm m})$ &$-$             &$0.1150$   &$0.1115$  &$-$            &$0.0860$  &$0.0795$   &$-$            &$0.0580$ &$0.0565$\\
			&$\sigma(H_0)$                      &$0.125$     &$0.101$     &$0.108$    &$0.120$    &$0.095$    &$0.100$      &$0.125$     &$0.086$   &$0.093$\\
			&$\sigma(w)$                          &$0.370$     &$0.205$    &$0.200$    &$0.355$   &$0.160$    &$0.160$      &$0.355$    &$0.120$   &$0.120$\\
			&$\ve(\Omega_{\rm m})$       &$-$            &$37.95\%$&$36.44\%$&$-$            &$28.29\%$&$26.15\%$&$-$             &$18.77\%$&$18.17\%$\\
			&$\ve(H_0)$                            &$0.187\%$&$0.151\%$ &$0.162\%$&$0.179\%$&$0.142\%$&$0.150\%$ &$0.187\%$&$0.129\%$&$0.139\%$\\
			&$\ve(w)$                                &$31.62\%$&$20.92\%$&$20.41\%$&$31.42\%$&$16.49\%$&$16.67\%$ &$31.42\%$&$12.50\%$&$12.50\%$\\
			\hline
			\multirow{6}{*}{ET2CE}
			&$\sigma(\Omega_{\rm m})$ &$-$              &$0.0690$ &$0.0650$  &$0.1750$    &$0.0520$  &$0.0440$ &$-$              &$0.0320$  &$0.0310$\\
			&$\sigma(H_0)$                      &$0.087$     &$0.060$   &$0.060$     &$0.079$    &$0.054$    &$0.054$     &$0.084$     &$0.047$     &$0.049$\\
			&$\sigma(w)$                          &$0.330$     &$0.130$    &$0.125$      &$0.290$    &$0.099$    &$0.086$     &$0.315$     &$0.066$     &$0.064$\\
			&$\ve(\Omega_{\rm m})$       &$-$              &$21.97\%$&$19.94\%$ &$54.69\%$&$16.40\%$&$13.79\%$ &$-$              &$9.94\%$   &$9.66\%$\\
			&$\ve(H_0)$                            &$0.130\%$&$0.090\%$&$0.090\%$ &$0.118\%$&$0.081\%$&$0.081\%$  &$0.126\%$ &$0.070\%$&$0.073\%$\\
			&$\ve(w)$                                &$30.00\%$&$13.40\%$&$12.63\%$  &$27.36\%$&$10.17\%$ &$8.84\%$   &$28.64\%$&$6.77\%$   &$6.58\%$\\
			\hline \hline
	\end{tabular}}
\end{table*}

In this subsection, we analyze the constraint results on cosmological parameters from GW multi-messenger observations with different EM detection scenarios. Firstly, we investigate the constraint results from GW-kilonova detections to determine the optimal GW-kilonova survey project. We consider the $\Lambda$CDM and $w$CDM models under different survey projects in the three bands. The detailed constraint results are shown in Tables~\ref{tab4} and~\ref{tab5}.\footnote{Note that GW-kilonova detections provide rather poor constraints on the $w_0w_a$CDM model and thus we only show the constraint results of the $\Lambda$CDM and $w$CDM models.} We find that, within the same band, LSST and CSST provide better constraints due to their higher numbers of detections among the three survey projects, whereas WFST gives the worst constraints as it has the lowest number of detections. In Fig.~\ref{fig7}, we show the constraint results in the $\Omega_{\rm m}$--$H_0$ plane for the $\Lambda$CDM model using GW-kilonova detections in the $i$ band under different survey projects, effectively reflecting the aforementioned conclusions. Meanwhile, we find that among the three bands across different survey projects, LSST in the $i$ band gives the best constraints since it has the highest number of detections. Quantitatively, in the $i$ band for the $\Lambda$CDM model, ET-LSST gives $\sigma(H_0)=0.047$ and $\sigma(\Omega_{\rm m})=0.0051$, ET2CE-LSST gives $\sigma(H_0)=0.027$ and $\sigma(\Omega_{\rm m})=0.0028$, as shown in Table~\ref{tab4}. CSST in the $i$ band gives similar constraints. Therefore, in the following discussions, we take LSST in the $i$ band as the representative survey project for GW-kilonova detections.

\begin{figure*}[htbp]
	\includegraphics[width=0.45\linewidth,angle=0]{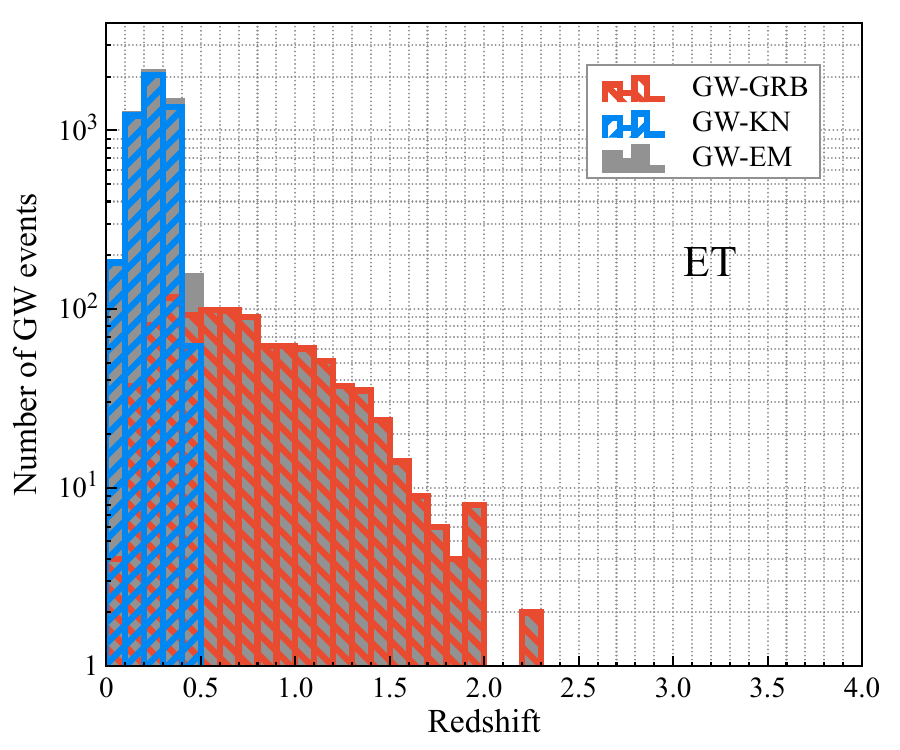}
	\includegraphics[width=0.45\linewidth,angle=0]{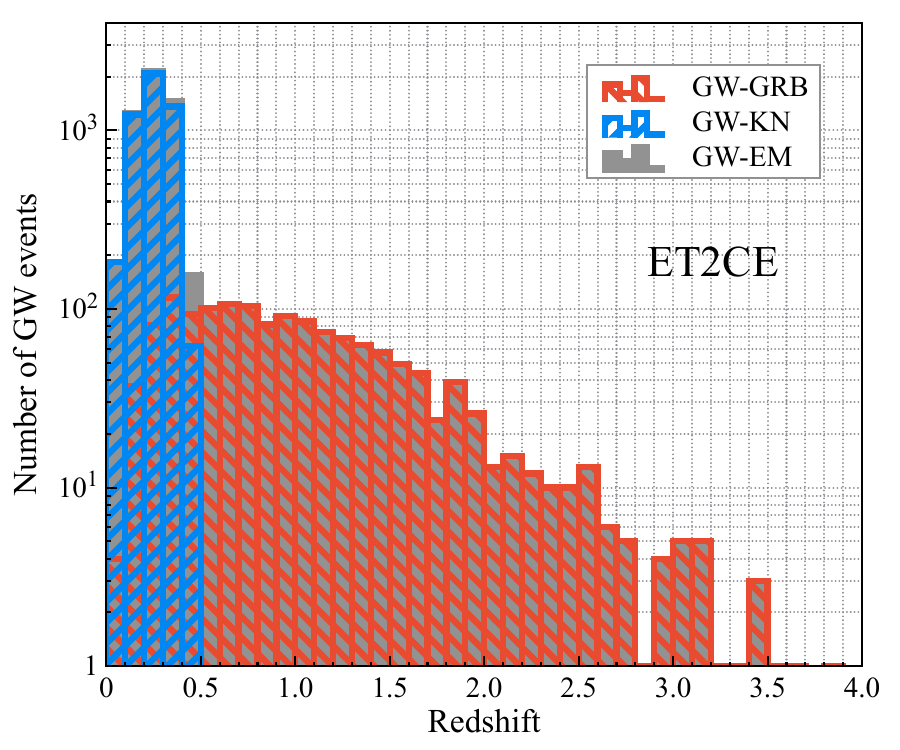}
	\caption{\label{fig8} Redshift distributions of GW-kilonova, GW-GRB, and GW-EM detections in a 10-year observation. The left panel shows the results of ET and the right panel shows those of ET2CE.}
\end{figure*}

\begin{table*}[!htb]
	\caption{Numbers of GW standard sirens in cosmological analysis detected by ET and ET2CE, with different EM detection scenarios in a 10-year observation.}
	\label{tab6}
	\setlength\tabcolsep{9.05mm}
	\renewcommand{\arraystretch}{2}
	\begin{tabular}{ccccc}
		\hline \hline
		Detection strategy&GW-KN&GW-GRB	&GW-KN\&GRB	&GW-EM\\
		\hline
		ET       &      4863      &   992    &    50     &  5805    \\
		ET2CE&      4917  &   1455   &    50   &    6322  \\
		\hline\hline
	\end{tabular}
\end{table*}

Then, we present the numbers of GW standard sirens used in cosmological analysis as detected by ET and ET2CE under various EM detection scenarios over a 10-year observation, as summarized in Table~\ref{tab6}. It should be noted that some sources can be detected by both GW-GRB and GW-kilonova detections. Therefore, the total number of GW-EM detections is not simply the sum of the GW-kilonova and GW-GRB detections. Specifically, in our analysis of GW-EM detections, we identify and remove duplicated events appearing in both channels, ensuring that each joint detection is counted only once when deriving event statistics and performing cosmological parameter estimation. In Fig.~\ref{fig8}, we present the redshift distributions of GW-kilonova, GW-GRB, and GW-EM detections detected by ET and ET2CE for a 10-year observation. We can clearly see that at redshifts below $\sim 0.4$, EM counterparts are primarily identified through kilonova observations, whereas at redshifts above $\sim 0.4$, GRBs play a dominant role in EM counterpart detections. Furthermore, the redshift distributions of GW-EM detections provided by ET and ET2CE are nearly identical for redshifts below $\sim 0.6$. This similarity can be attributed to the fact that the detection efficiencies of ET and ET2CE in the low-redshift regime are not substantially different, as mentioned earlier. As a result, within this range, the choice between a single GW detector or their network has little impact on EM counterpart detections. However, as the redshift increases, ET2CE demonstrates a greater capability to detect GW events at deeper redshifts compared to ET, as illustrated in Fig.~\ref{fig1}. Consequently, the EM counterparts associated with ET2CE can also be identified at significantly higher redshifts than those associated with ET.

\begin{figure*}[htbp]
	\includegraphics[width=0.4\linewidth,angle=0]{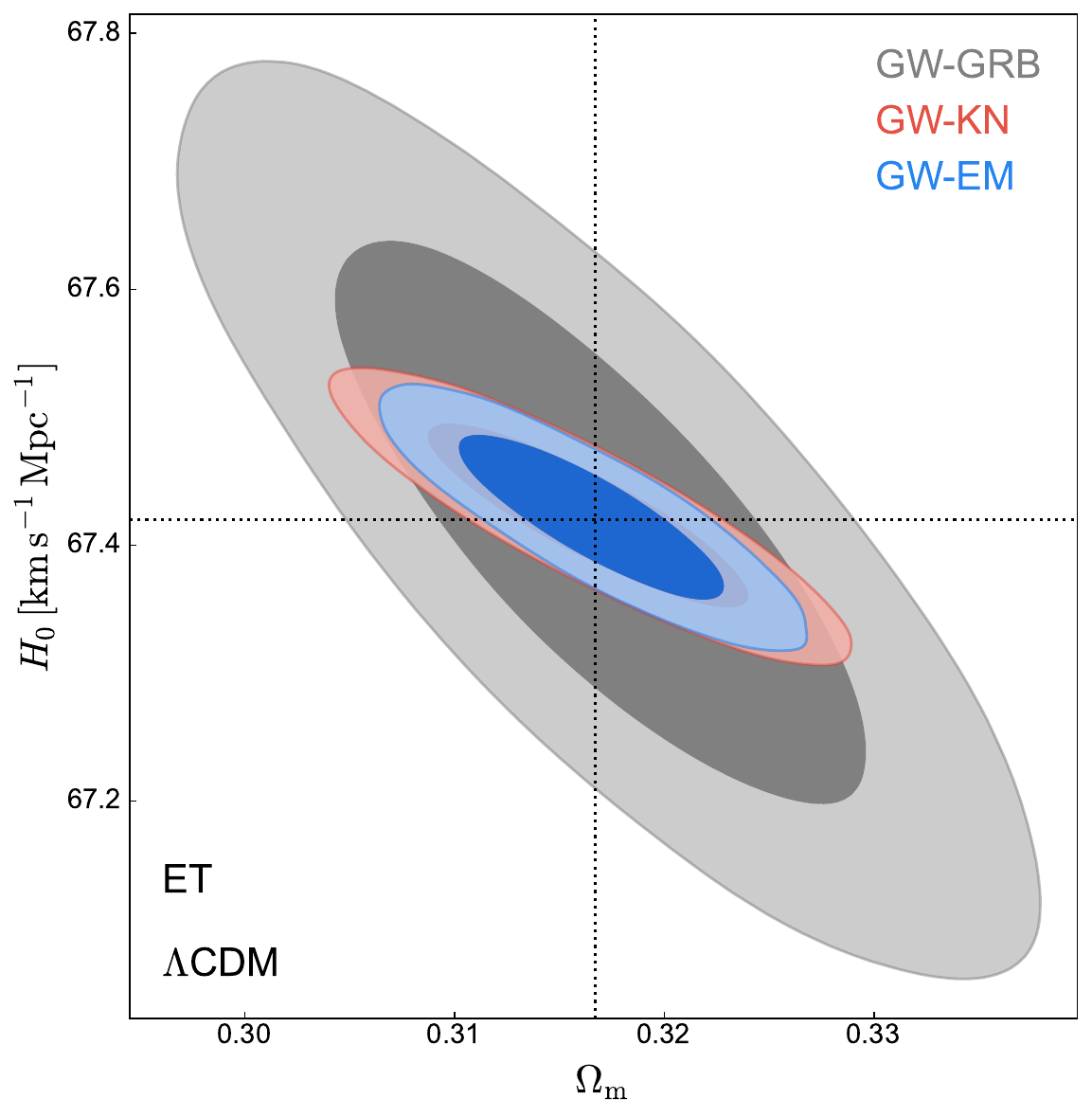}
	\includegraphics[width=0.4\linewidth,angle=0]{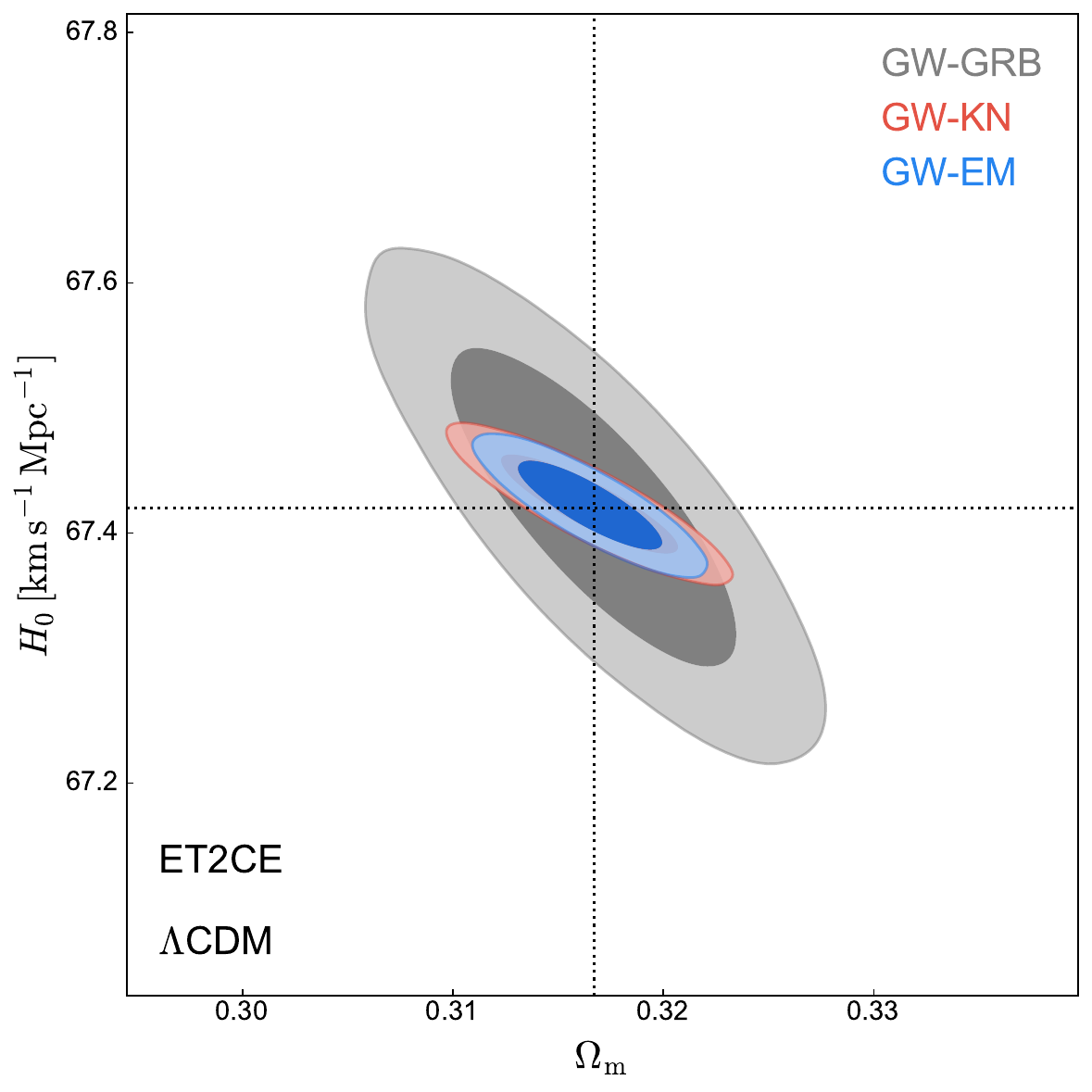}
	\caption{\label{fig9} Two-dimensional marginalized contours ($68.3\%$ and $95.4\%$ confidence level) in the $\Omega_{\rm m}$--$H_0$ plane for the $\Lambda$CDM model using GW-kilonova, GW-GRB, and GW-EM detections. The left panel shows the constraint results from ET and the right panel shows those from ET2CE. Here, the dotted lines indicate the fiducial values of cosmological parameters preset in the simulation.}
\end{figure*}

\begin{figure*}[htbp]
	\includegraphics[width=0.4\linewidth,angle=0]{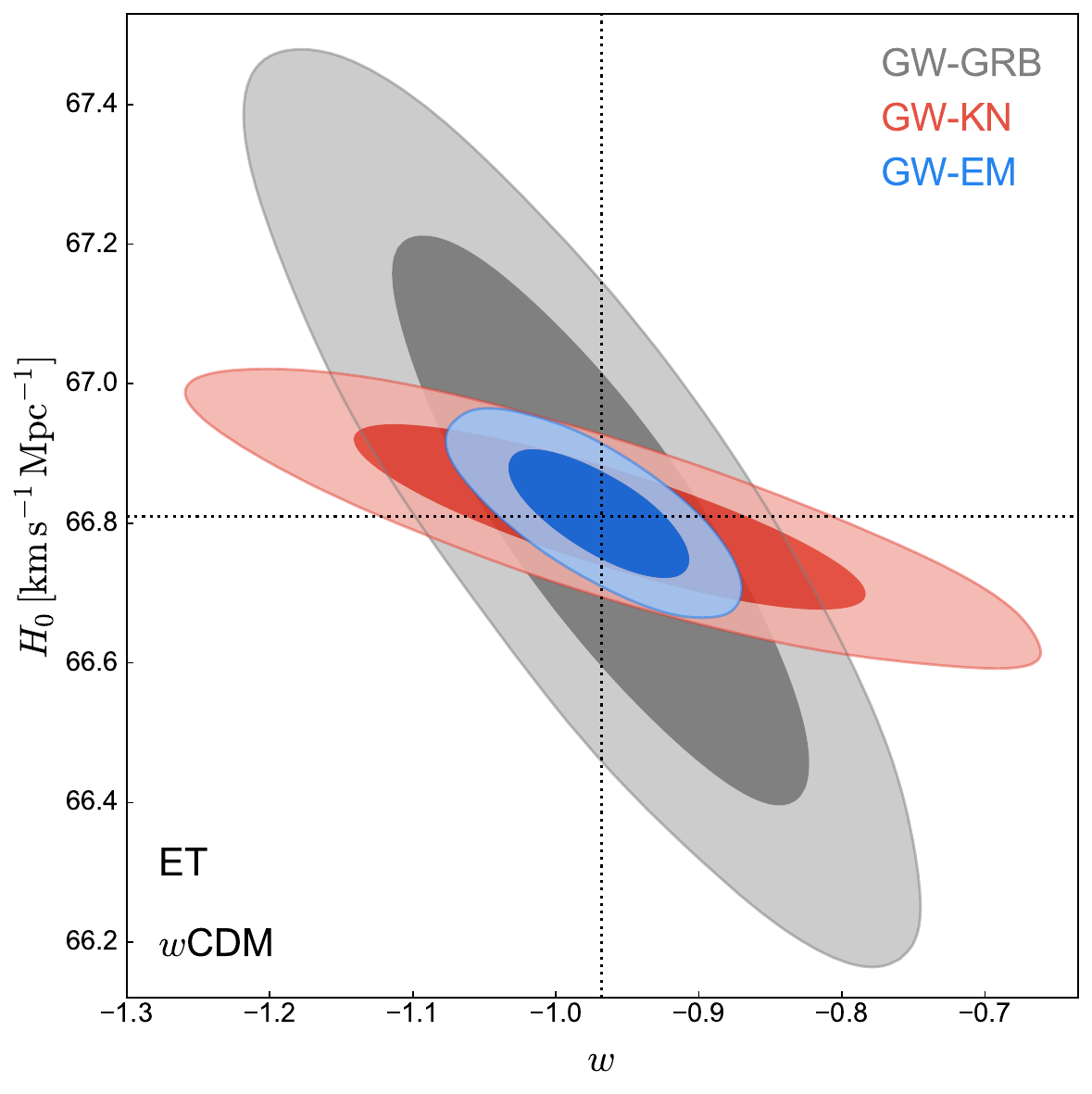}
	\includegraphics[width=0.4\linewidth,angle=0]{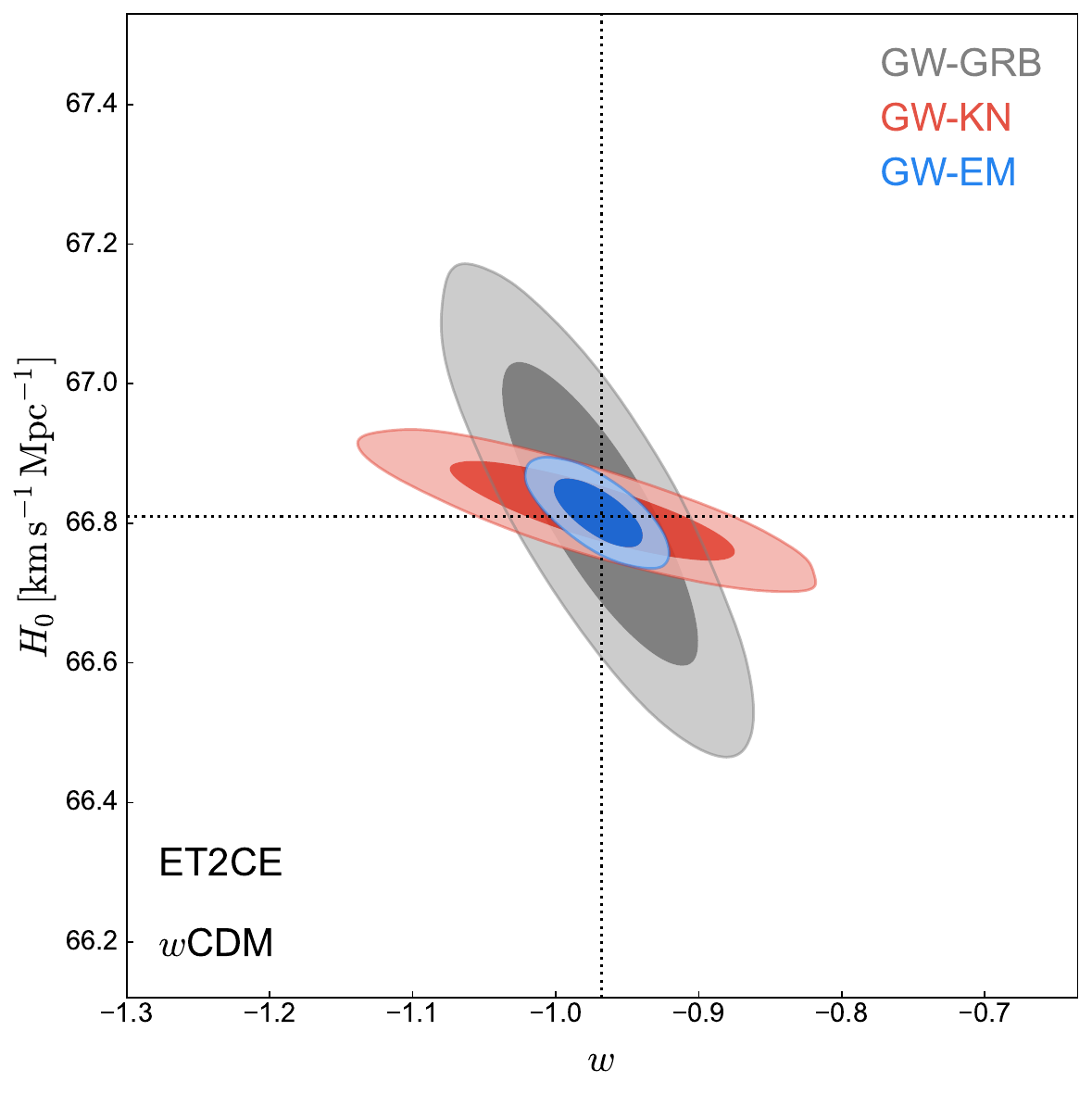}
	\includegraphics[width=0.4\linewidth,angle=0]{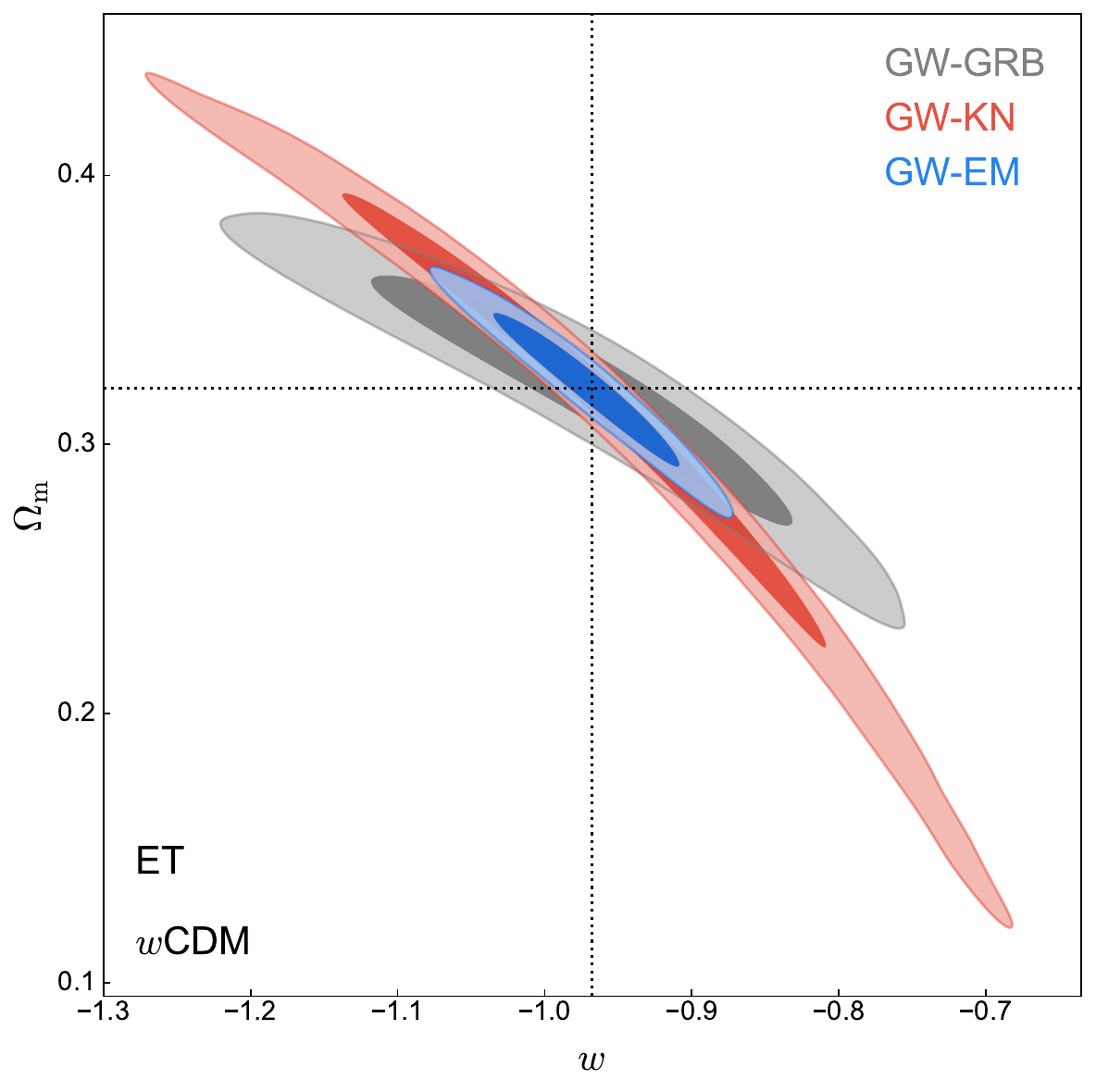}
	\includegraphics[width=0.4\linewidth,angle=0]{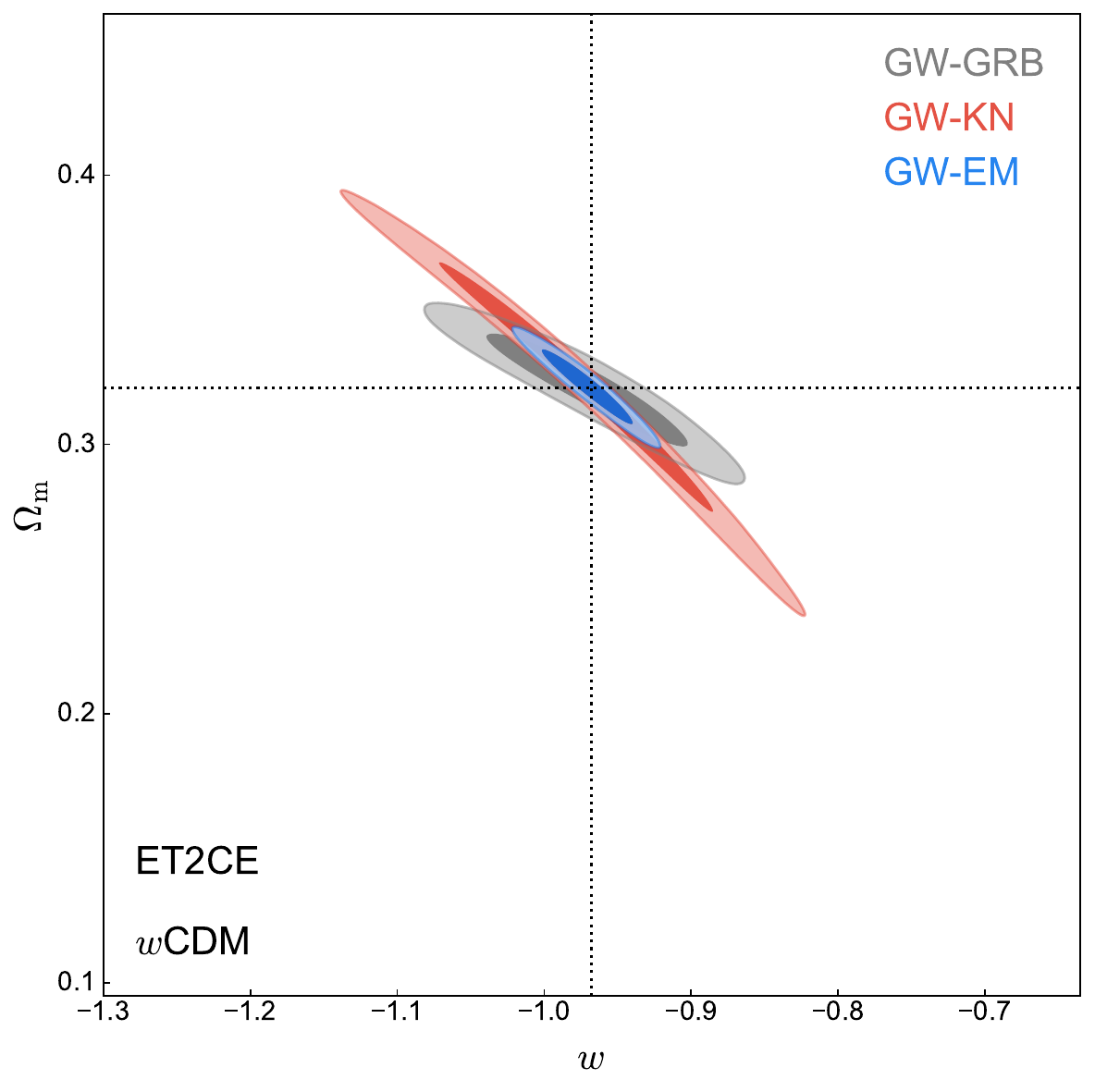}
	\caption{\label{fig10} Same as Fig.~\ref{fig9}, but in the $w$--$H_0$ and $w$--$\Omega_{\rm m}$ planes for the $w$CDM model.}
\end{figure*}

\begin{figure*}[htbp]
	\includegraphics[width=0.4\linewidth,angle=0]{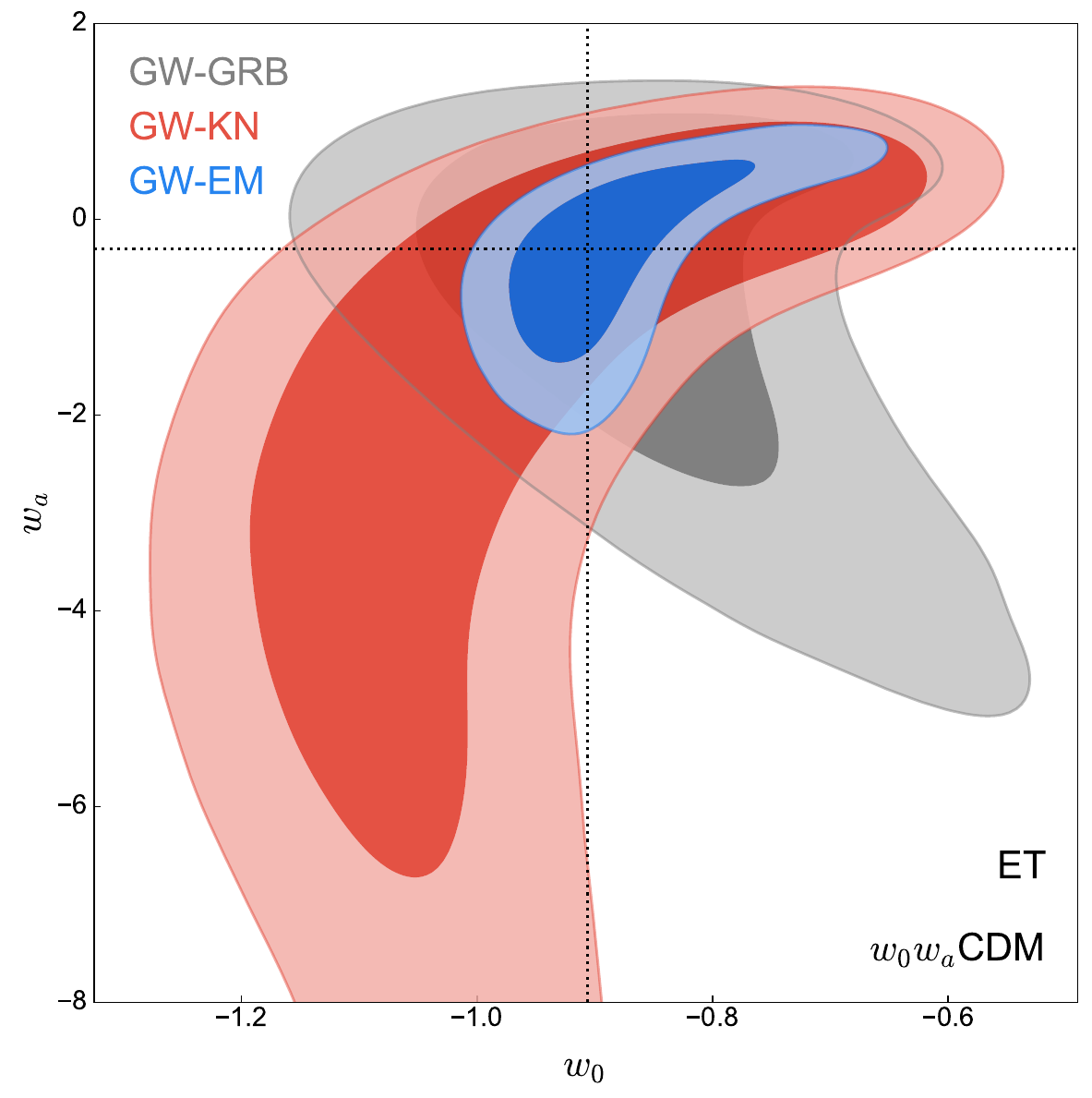}
	\includegraphics[width=0.4\linewidth,angle=0]{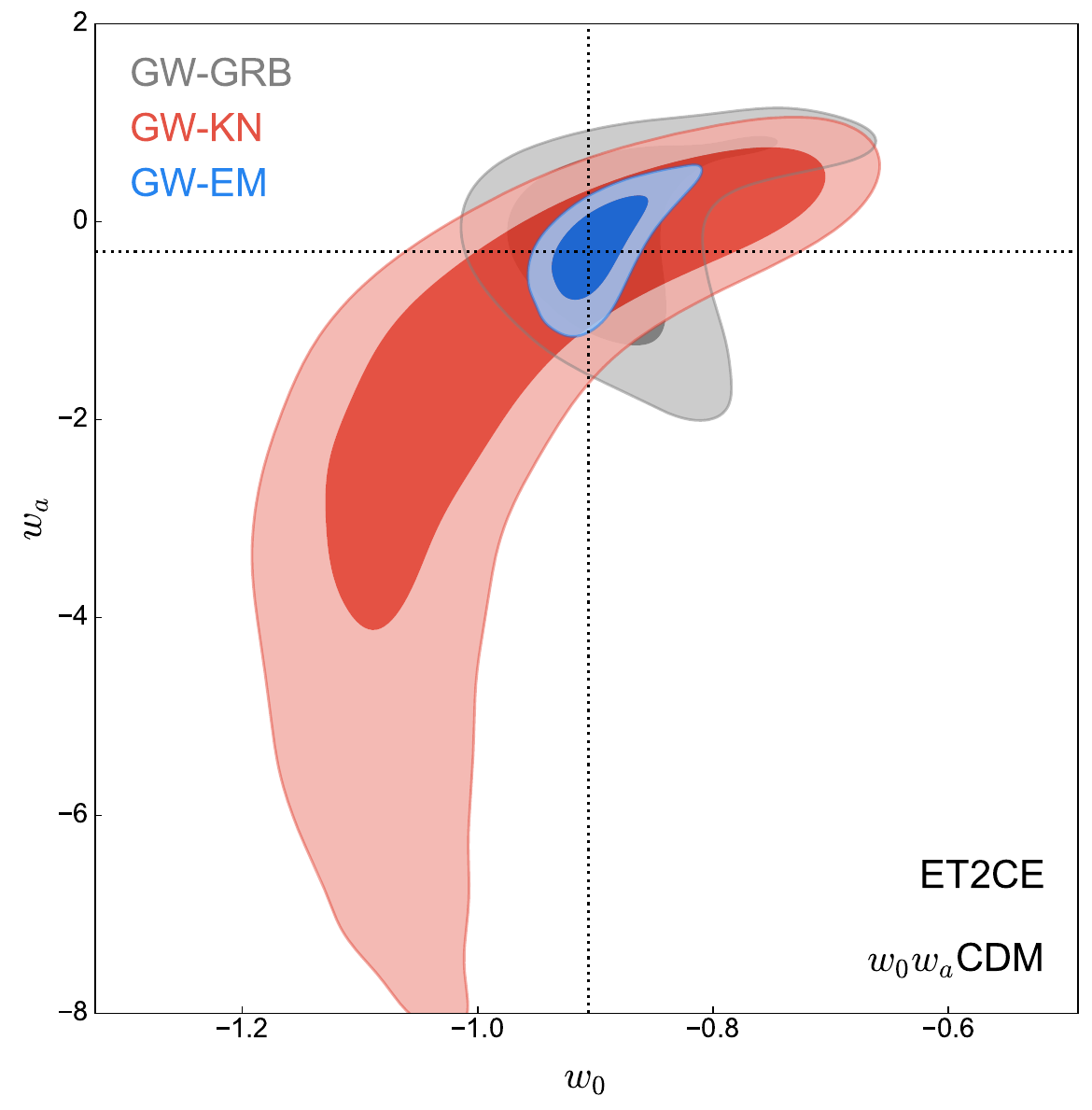}
	\caption{\label{fig11} Same as Fig.~\ref{fig9}, but in the $w_0$--$w_a$ plane for the $w_0w_a$CDM model.}
\end{figure*}

\begin{table*}[!htb]
	\caption{ Absolute (1$\sigma$) and relative errors of cosmological parameters in the $\Lambda$CDM model using the GW-GRB, GW-KN, GW-EM, CBS, CBS+GW-GRB, CBS+GW-KN, and CBS+GW-EM data under the GW detection strategies of ET and ET2CE. Here, $H_0$ is in units of $\rm km\ s^{-1}\ Mpc^{-1}$.}
	\label{tab7}
	\setlength{\tabcolsep}{2.75mm}
	\renewcommand{\arraystretch}{1.5}
	\begin{center}{\centerline{
				\begin{tabular}{c|c|m{1.5cm}<{\centering}|m{1.5cm}<{\centering}|m{1.5cm}<{\centering}|m{1.5cm}<{\centering}|m{1.5cm}<{\centering}|m{1.5cm}<{\centering}|m{1.5cm}<{\centering}}
					\hline \hline
					\multirow{2}{*}{\shortstack{Detection \\ strategy}} &\multirow{2}{*}{Error} &\multirow{2}{*}{GW-GRB}   & \multirow{2}{*}{GW-KN} & \multirow{2}{*}{GW-EM} & \multirow{2}{*}{CBS}&\multirow{2}{*}{\shortstack{CBS+ \\ GW-GRB}}&\multirow{2}{*}{\shortstack{CBS+ \\ GW-KN}}& \multirow{2}{*}{\shortstack{CBS+ \\ GW-EM}} \\ &&&&&&&&\\
					\hline
					\multirow{4}{*}{ET}
					&$\sigma(\Omega_{\rm m})$       &$0.0084$   &$0.0051$    &$0.0042$    &$0.0059$     &$0.0017$    &$0.0011$&$0.0010$   \\
					&$\sigma(H_0)$                            &$0.150$       &$0.047$       &$0.043$    &$0.420$        &$0.100$     &$0.027$&$0.026$    \\
					&$\ve(\Omega_{\rm m})$             &$2.65\%$  &$1.61\%$      &$1.33\%$     &$1.86\%$     &$0.54\%$ &$0.35\%$&$0.32\%$ \\
					&$\ve(H_0)$                                  &$0.222\%$ &$0.070\%$  &$0.064\%$ &$0.623\%$   &$0.148\%$   &$0.040\%$&$0.039\%$  \\
					\hline
					\multirow{4}{*}{ET2CE}
					&$\sigma(\Omega_{\rm m})$       &$0.0045$   &$0.0028$    &$0.0023$    &$0.0059$     &$0.0013$    &$0.0010$&$0.0010$   \\
					&$\sigma(H_0)$                            &$0.084$     &$0.027$      &$0.023$       &$0.420$         &$0.063$     &$0.016$&$0.016$    \\
					&$\ve(\Omega_{\rm m})$              &$1.42\%$  &$0.88\%$     &$0.73\%$    &$1.86\%$      &$0.41\%$     &$0.32\%$&$0.32\%$ \\
					&$\ve(H_0)$                                  &$0.125\%$ &$0.040\%$  &$0.034\%$  &$0.623\%$   &$0.093\%$   &$0.024\%$&$0.024\%$  \\
					\hline
					\hline
		\end{tabular}}}
	\end{center}
\end{table*}

\begin{table*}[!htb]
	\caption{ Same as Table~\ref{tab7}, but in the $w$CDM model.}
	\label{tab8}
	\setlength{\tabcolsep}{2.75mm}
	\renewcommand{\arraystretch}{1.5}
	\begin{center}{\centerline{
				\begin{tabular}{c|c|m{1.5cm}<{\centering}|m{1.5cm}<{\centering}|m{1.5cm}<{\centering}|m{1.5cm}<{\centering}|m{1.5cm}<{\centering}|m{1.5cm}<{\centering}|m{1.5cm}<{\centering}}
					\hline \hline
					\multirow{2}{*}{\shortstack{Detection \\ strategy}} &\multirow{2}{*}{Error} &\multirow{2}{*}{GW-GRB}   & \multirow{2}{*}{GW-KN} & \multirow{2}{*}{GW-EM} & \multirow{2}{*}{CBS}&\multirow{2}{*}{\shortstack{CBS+ \\ GW-GRB}}&\multirow{2}{*}{\shortstack{CBS+ \\ GW-KN}}& \multirow{2}{*}{\shortstack{CBS+ \\ GW-EM}} \\ &&&&&&&&\\
					\hline
					\multirow{6}{*}{ET}
					&$\sigma(\Omega_{\rm m})$       &$0.0310$   &$0.0580$    &$0.0190$    &$0.0070$     &$0.0020$   &$0.0015$   &$0.0015$   \\
					&$\sigma(H_0)$                            &$0.270$      &$0.086$     &$0.061$      &$0.690$        &$0.140$     &$0.048$     &$0.046$    \\
					&$\sigma(w)$                                &$0.095$    &$0.120$        &$0.042$     &$0.028$        &$0.017$      &$0.011$      &$0.010$   \\
					&$\ve(\Omega_{\rm m})$              &$9.69\%$  &$18.77\%$  &$5.92\%$   &$2.18\%$      &$0.62\%$    &$0.47\%$   &$0.47\%$ \\
					&$\ve(H_0)$                                  &$0.404\%$ &$0.129\%$  &$0.091\%$ &$1.033\%$   &$0.210\%$   &$0.072\%$&$0.069\%$  \\
					&$\ve(w)$                                        &$9.76\%$ &$12.50\%$  &$4.33\%$    &$2.89\%$    &$1.75\%$     &$1.14\%$    &$1.03\%$  \\ \hline			
					\multirow{6}{*}{ET2CE}
					&$\sigma(\Omega_{\rm m})$       &$0.0135$   &$0.0320$    &$0.0091$       &$0.0070$     &$0.0017$    &$0.0013$&$0.0013$   \\
					&$\sigma(H_0)$                            &$0.140$      &$0.047$     &$0.033$         &$0.690$       &$0.098$     &$0.028$&$0.027$    \\
					&$\sigma(w)$                                 &$0.045$    &$0.066$  &$0.020$            &$0.028$        &$0.014$     &$0.007$&$0.007$   \\
					&$\ve(\Omega_{\rm m})$              &$4.21\%$  &$9.94\%$  &$2.83\%$        &$2.18\%$       &$0.53\%$  &$0.41\%$&$0.41\%$ \\
					&$\ve(H_0)$                                  &$0.210\%$ &$0.070\%$  &$0.049\%$   &$1.033\%$    &$0.147\%$  &$0.042\%$&$0.040\%$  \\
					&$\ve(w)$                                       &$4.64\%$ &$6.77\%$  &$2.06\%$         &$2.89\%$      &$1.45\%$   &$0.72\%$&$0.72\%$  \\
					\hline
					\hline
		\end{tabular}}}
	\end{center}
\end{table*}

\begin{table*}[!htb]
	\caption{ Same as Table~\ref{tab7}, but in the $w_0w_a$CDM model.}
	\label{tab9}
	\setlength{\tabcolsep}{2.75mm}
	\renewcommand{\arraystretch}{1.5}
	\begin{center}{\centerline{
				\begin{tabular}{c|c|m{1.5cm}<{\centering}|m{1.5cm}<{\centering}|m{1.5cm}<{\centering}|m{1.5cm}<{\centering}|m{1.5cm}<{\centering}|m{1.5cm}<{\centering}|m{1.5cm}<{\centering}}
					\hline \hline
					\multirow{2}{*}{\shortstack{Detection \\ strategy}} &\multirow{2}{*}{Error} &\multirow{2}{*}{GW-GRB}   & \multirow{2}{*}{GW-KN} & \multirow{2}{*}{GW-EM} & \multirow{2}{*}{CBS}&\multirow{2}{*}{\shortstack{CBS+ \\ GW-GRB}}&\multirow{2}{*}{\shortstack{CBS+ \\ GW-KN}}& \multirow{2}{*}{\shortstack{CBS+ \\ GW-EM}} \\ &&&&&&&&\\
					\hline
					\multirow{7}{*}{ET}
					&$\sigma(\Omega_{\rm m})$       &$0.0760$   &$0.1245$    &$0.0620$     &$0.0070$     &$0.0039$    &$0.0030$  &$0.0029$   \\
					&$\sigma(H_0)$                            &$0.465$     &$0.165$     &$0.094$         &$0.700$       &$0.240$       &$0.060$   &$0.060$    \\
					&$\sigma(w_0)$                           &$0.110$       &$0.145$      &$0.056$        &$0.068$       &$0.042$     &$0.015$     &$0.014$   \\
					&$\sigma(w_a)$                           &$1.22$         &$2.30$        &$0.70$          &$0.29$         &$0.18$         &$0.11$        &$0.11$   \\
					&$\ve(\Omega_{\rm m})$             &$24.28\%$ &$31.76\%$  &$20.60\%$  &$2.18\%$      &$1.22\%$     &$0.94\%$   &$0.91\%$ \\
					&$\ve(H_0)$                                  &$0.695\%$ &$0.247\%$  &$0.140\%$  &$1.045\%$   &$0.358\%$   &$0.090\%$&$0.090\%$  \\
					&$\ve(w_0)$                                  &$12.94\%$ &$14.95\%$  &$6.33\%$    &$7.51\%$      &$4.63\%$     &$1.65\%$    &$1.54\%$  \\ \hline
					\multirow{7}{*}{ET2CE}
					&$\sigma(\Omega_{\rm m})$       &$0.0485$   &$0.1190$      &$0.0290$     &$0.0070$    &$0.0034$    &$0.0029$&$0.0027$   \\
					&$\sigma(H_0)$                            &$0.230$     &$0.108$        &$0.051$        &$0.700$      &$0.150$     &$0.036$    &$0.036$    \\
					&$\sigma(w_0)$                           &$0.054$      &$0.123$        &$0.025$       &$0.068$      &$0.028$    & $0.009$   &$0.009$   \\
					&$\sigma(w_a)$                           &$0.70$         &$1.73$          &$0.35$         &$0.29$         &$0.14$       &$0.10$      &$0.09$   \\
					&$\ve(\Omega_{\rm m})$              &$16.67\%$ &$30.83\%$  &$9.32\%$     &$2.18\%$     &$1.06\%$   &$0.91\%$  &$0.84\%$ \\
					&$\ve(H_0)$                                  &$0.343\%$ &$0.161\%$  &$0.076\%$    &$1.045\%$   &$0.224\%$&$0.054\%$&$0.054\%$  \\
					&$\ve(w_0)$                                  &$6.14\%$    &$12.63\%$  &$2.78\%$      &$7.51\%$      &$3.09\%$  &$0.99\%$  &$0.99\%$  \\
					\hline
					\hline
		\end{tabular}}}
	\end{center}
\end{table*}

Subsequently, we examine the impact of different GW detection strategies on cosmological analysis under the same EM detection scenarios. In Figs.~\ref{fig9}--\ref{fig11}, we show the constraint results in the $\Lambda$CDM, $w$CDM, and $w_0w_a$CDM models using GW-kilonova, GW-GRB, and GW-EM detections. The detailed results are given in Tables~\ref{tab7}--\ref{tab9}. We can see that under the same EM detection scenarios, ET2CE gives better constraints on cosmological parameters compared to ET. For the GW-GRB detections, ET2CE gives $\sigma(\Omega_{\rm m})=0.0045$ and $\sigma(H_0)=0.084$ in the $\Lambda$CDM model, which are $46.4\%$ and $44.0\%$ better than those from ET. For the constraint on $w$ in the $w$CDM model, ET2CE gives $\sigma(w)=0.045$ with $52.6\%$ better than that from ET. For the parameter $w_0$ in the $w_0w_a$CDM model, ET2CE gives $\sigma(w_0)=0.054$ with $50.9\%$ better than that of ET. For the GW-kilonova detections, ET2CE provides $\sigma(H_0)=0.027$, $\sigma(w)=0.066$, and $\sigma(w_0)=0.123$ in the $\Lambda$CDM, $w$CDM, and $w_0w_a$CDM models, respectively, which are $42.6\%$, $45.0\%$, and $15.2\%$ better than those from ET. In the case of GW-EM detections, ET2CE provides $\sigma(H_0)=0.023$, $\sigma(w)=0.020$, and $\sigma(w_0)=0.025$ in the $\Lambda$CDM, $w$CDM, and $w_0w_a$CDM models, respectively, which are $46.5\%$, $52.4\%$, and $55.4\%$ better than those from ET. As a result, in the following, we take ET2CE as the representative GW detection strategy for subsequent discussions.

Finally, we explore the effect of different EM detection scenarios on cosmological analysis under the same GW detection strategy. From Fig.~\ref{fig9}, we can find that under the same GW detection strategy, the constraints from GW-EM are better than those from GW-GRB and are slightly better than those from GW-kilonova in the $\Lambda$CDM model. Taking ET2CE as an example, GW-EM gives $\sigma(H_0)=0.023$, which is $72.6\%$ and $14.8\%$ better than those from GW-GRB and GW-kilonova. For the $w$CDM and $w_0w_a$CDM models shown in Figs.~\ref{fig10} and~\ref{fig11}, we find that, compared to GW-GRB, GW-kilonova provides weaker constraints on $\Omega_{\rm m}$ and the dark energy equation-of-state (EoS) parameters. However, GW-EM improves the constraints on these parameters compared to GW-GRB and GW-kilonova. For instance, with ET2CE, in the $w$CDM model, GW-EM provides $\sigma(w)=0.020$, which is $55.6\%$ and $69.7\%$ better than those from GW-GRB and GW-kilonova. For $w_0$ in the $w_0w_a$CDM model, GW-EM provides $\sigma(w_0)=0.025$, which is $53.7\%$ and $79.7\%$ better than those from GW-GRB and GW-kilonova.

In addition, from Tables~\ref{tab7}--\ref{tab9}, we can find that GW-kilonova significantly improves the constraints on $H_0$ compared to GW-GRB. This enhancement is not only a consequence of the larger number of detections, but also reflects differences in the redshift distributions of the EM detection scenarios in our simulations. As shown in Fig.~\ref{fig8}, the redshift distribution of GW-kilonova lies below $\sim 0.4$, a range where $H_0$ is most directly determined from the slope of the distance-redshift relation with minimal dependence on other cosmological parameters. In contrast, GW-GRB is subject to strong viewing-angle selection, resulting in far fewer low-redshift detections and a distribution shifted toward higher redshifts, which is less sensitive to $H_0$ and more affected by degeneracies with other parameters.

\subsection{Constraint results from GW multi-messenger observations combined with CBS observations}

\begin{figure*}[htbp]
	\includegraphics[width=0.4\linewidth,angle=0]{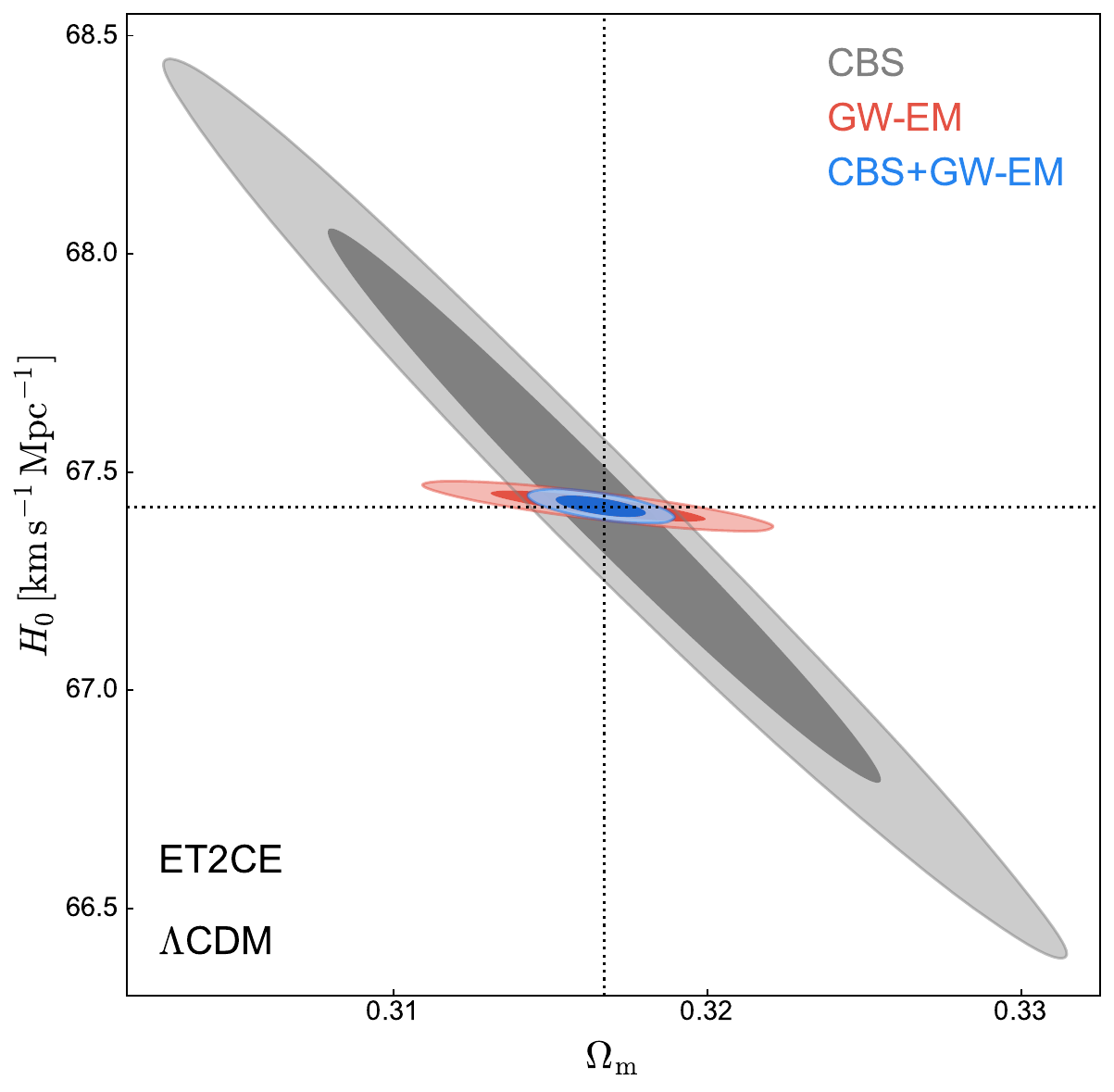}
	\includegraphics[width=0.4\linewidth,angle=0]{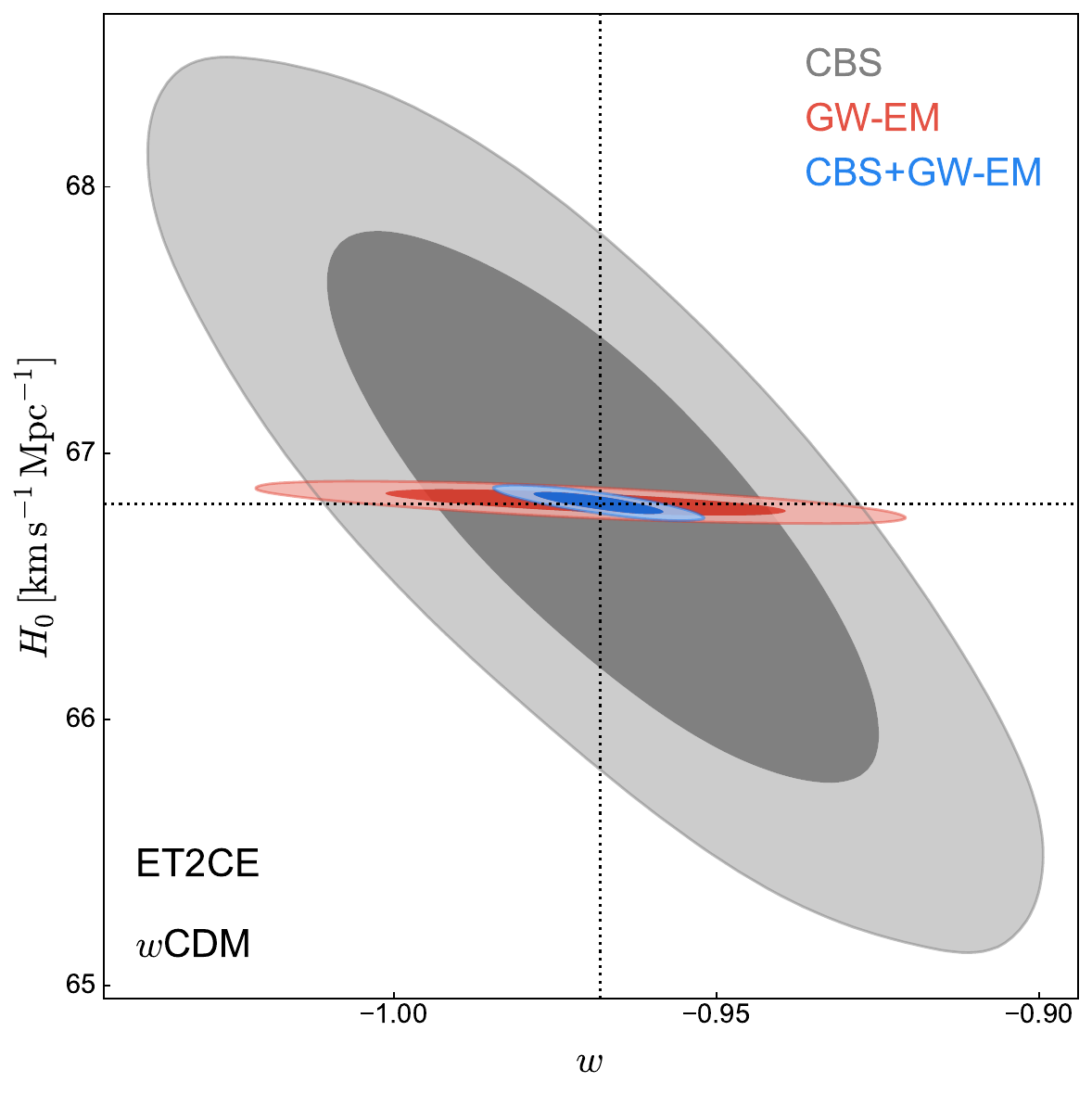}
	\includegraphics[width=0.4\linewidth,angle=0]{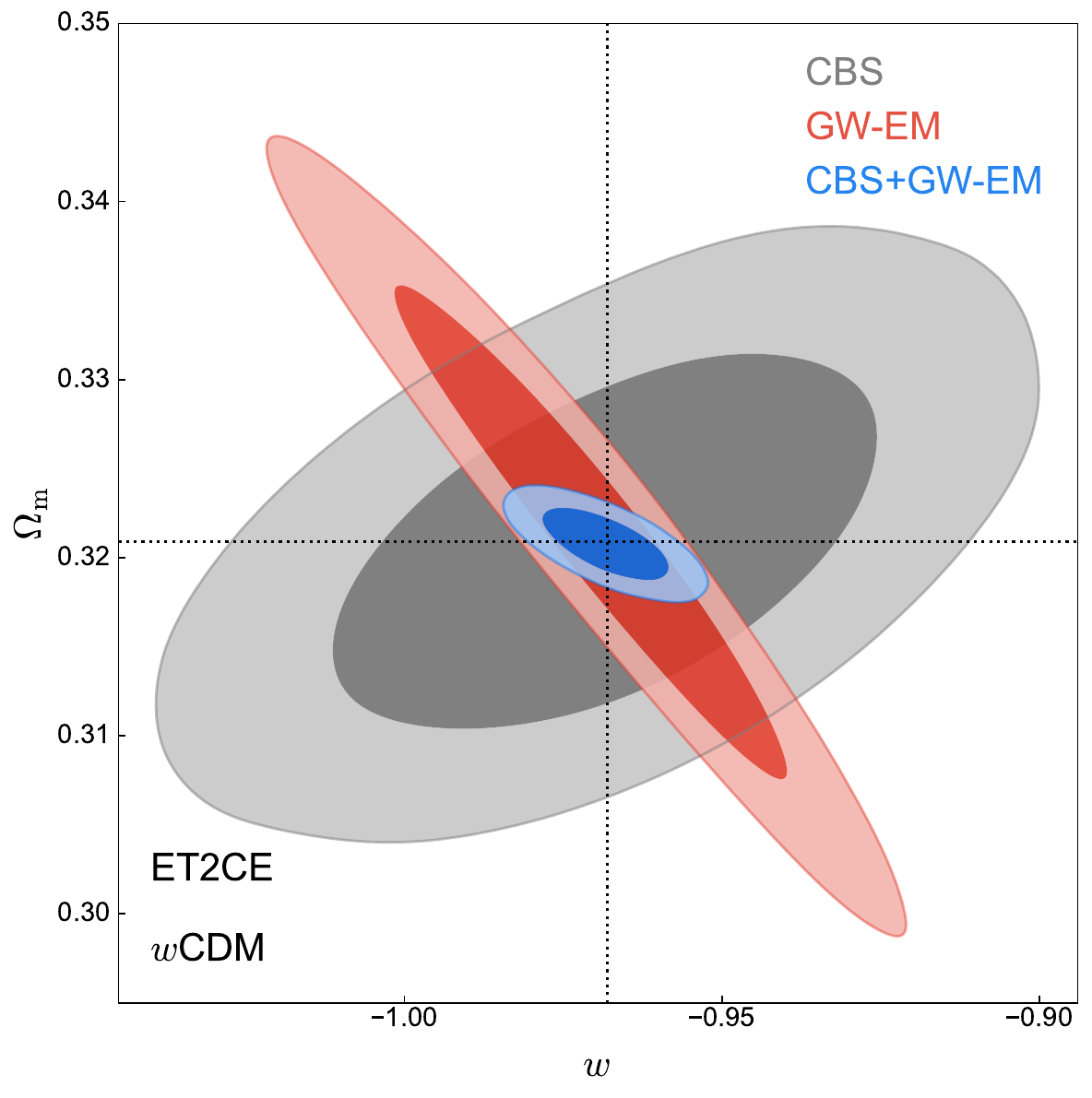}
	\includegraphics[width=0.4\linewidth,angle=0]{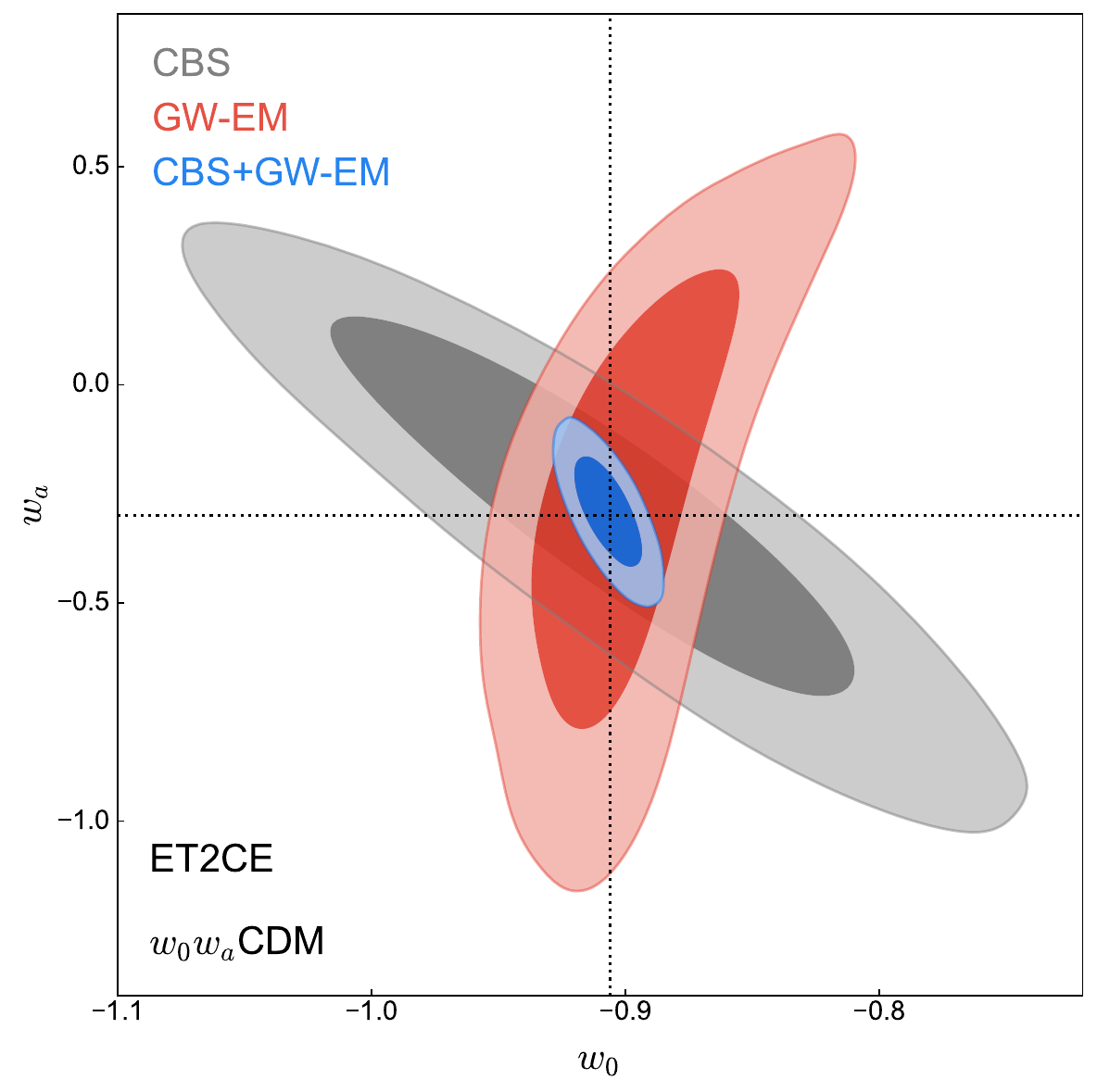}
	\caption{\label{fig12} Two-dimensional marginalized contours ($68.3\%$ and $95.4\%$ confidence level) in the $\Omega_{\rm m}$--$H_0$, $w$--$H_0$, $w$--$\Omega_{\rm m}$, and $w_0$--$w_a$ planes using the CBS, GW-EM, and CBS+GW-EM data in the $\Lambda$CDM, $w$CDM, and $w_0w_a$CDM models. Here, the GW data are simulated based on the GW detection strategy of ET2CE, and the dotted lines indicate the fiducial values of cosmological parameters preset in the simulation.}
\end{figure*}

\begin{figure*}[htbp]
	\includegraphics[width=0.4\linewidth,angle=0]{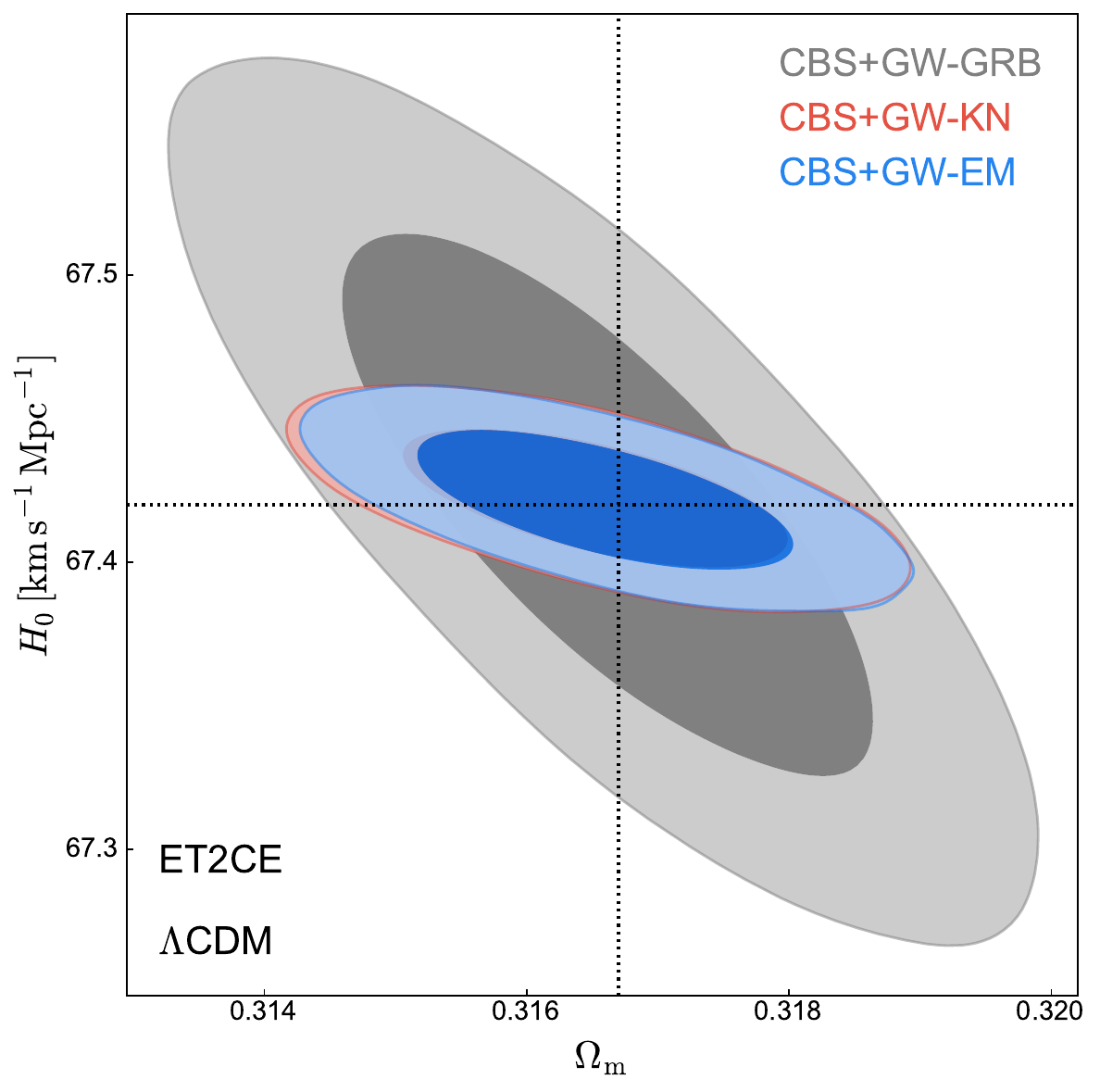}
	\includegraphics[width=0.4\linewidth,angle=0]{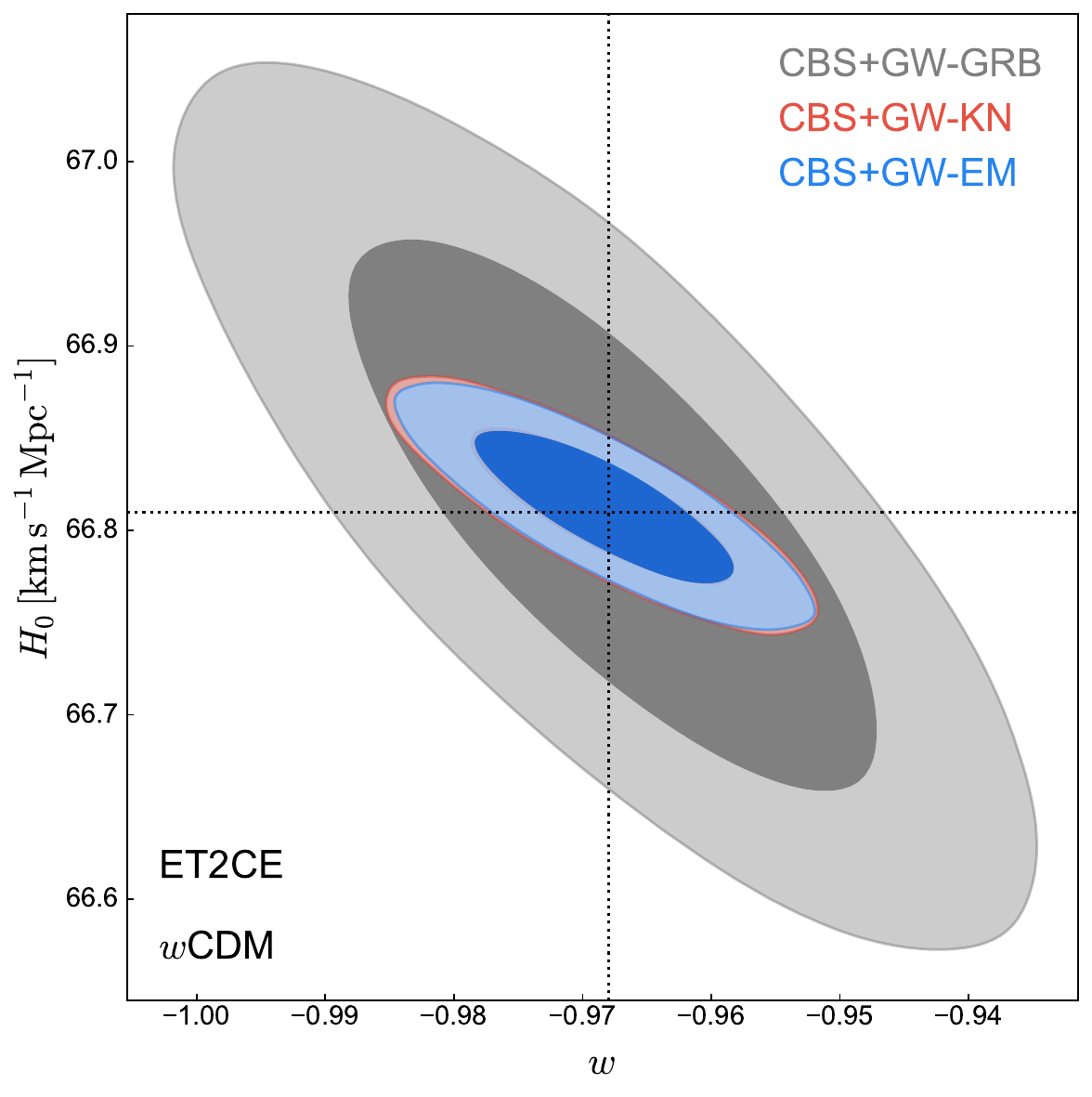}
	\includegraphics[width=0.4\linewidth,angle=0]{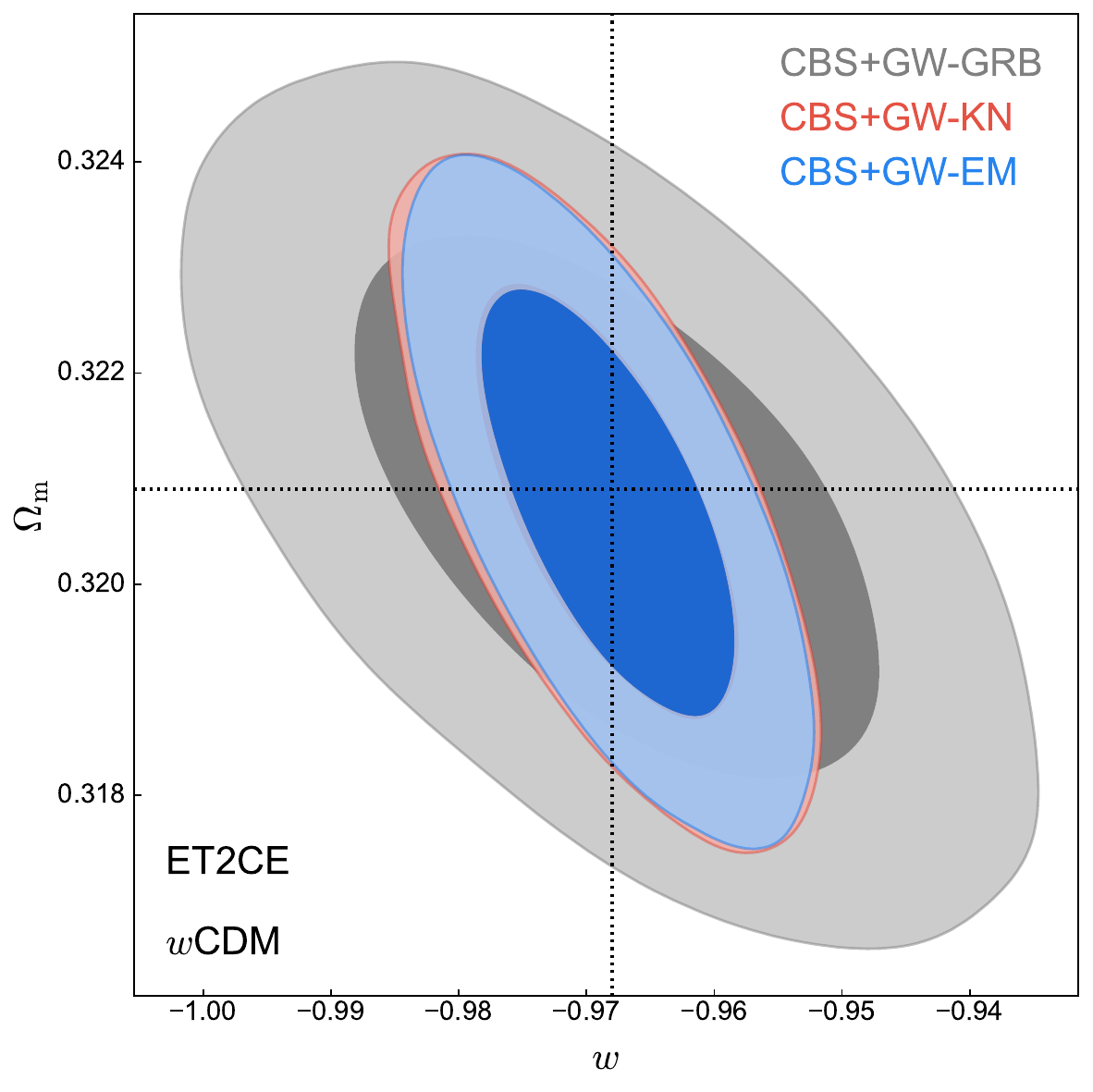}
	\includegraphics[width=0.4\linewidth,angle=0]{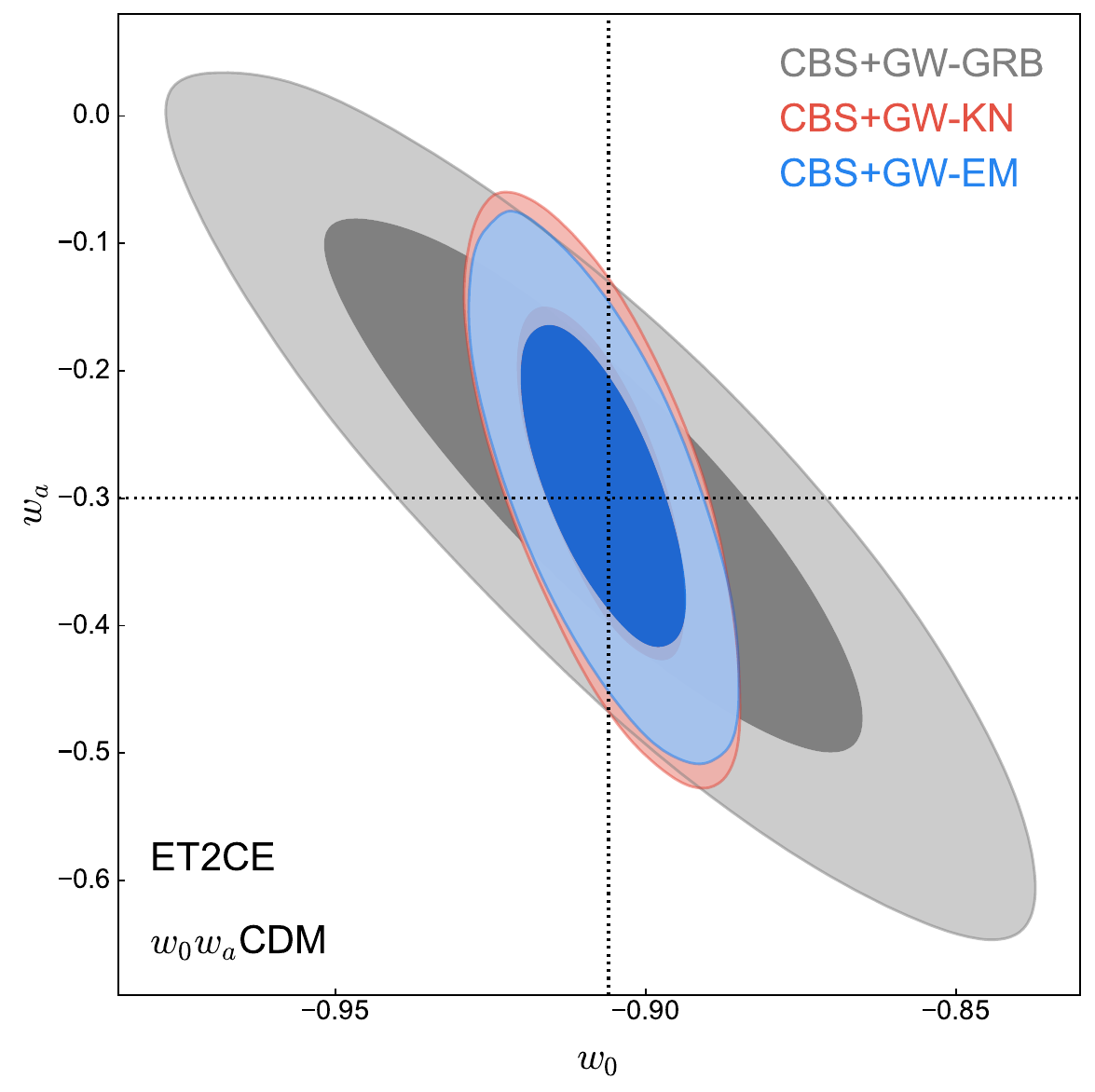}
	\caption{\label{fig13} Same as Fig~\ref{fig12}, but using the CBS+GW-GRB, CBS+GW-KN, and CBS+GW-EM data.}
\end{figure*}

\begin{figure*}[htbp]
	\includegraphics[width=0.4\linewidth,angle=0]{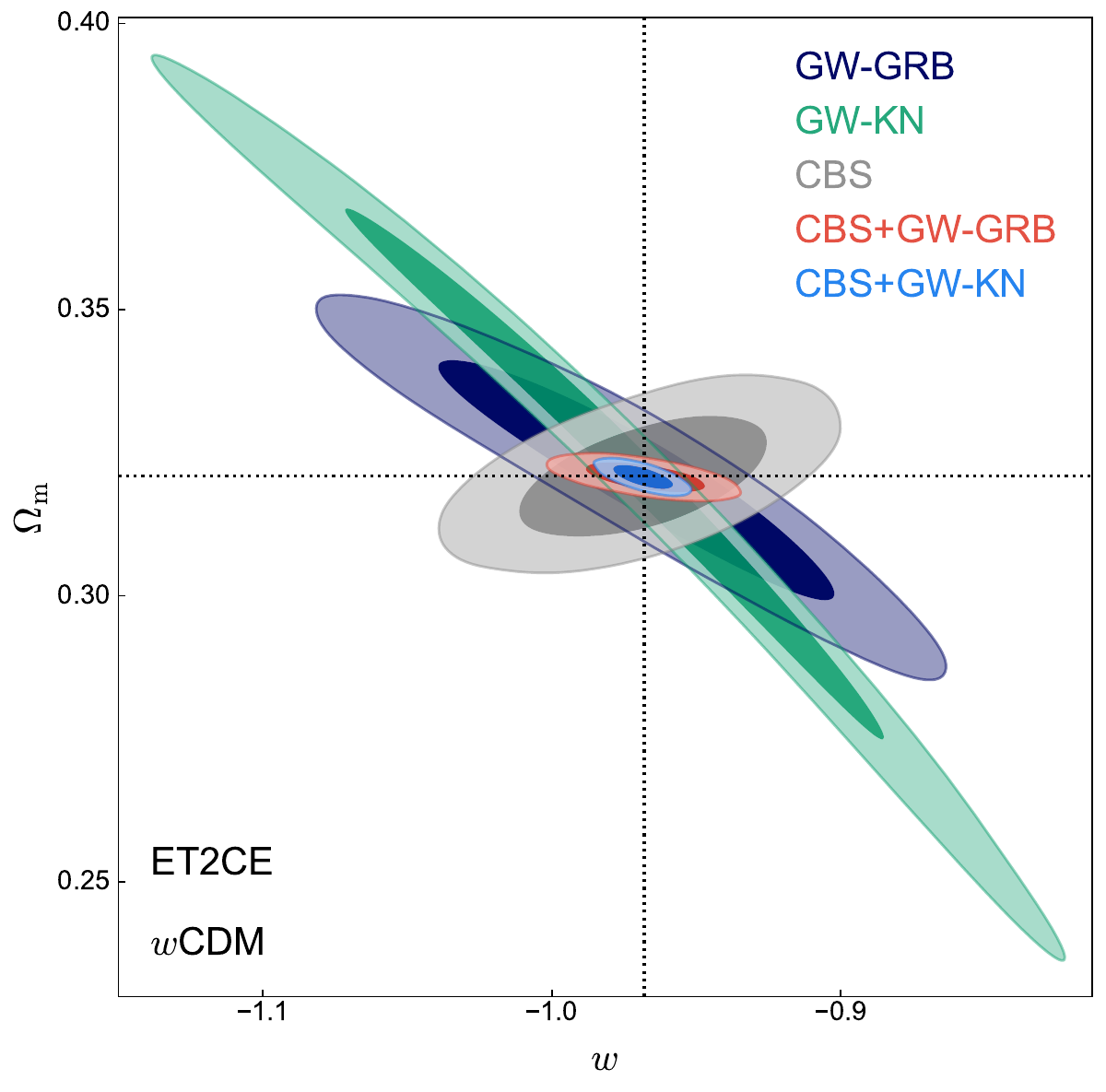}
	\includegraphics[width=0.4\linewidth,angle=0]{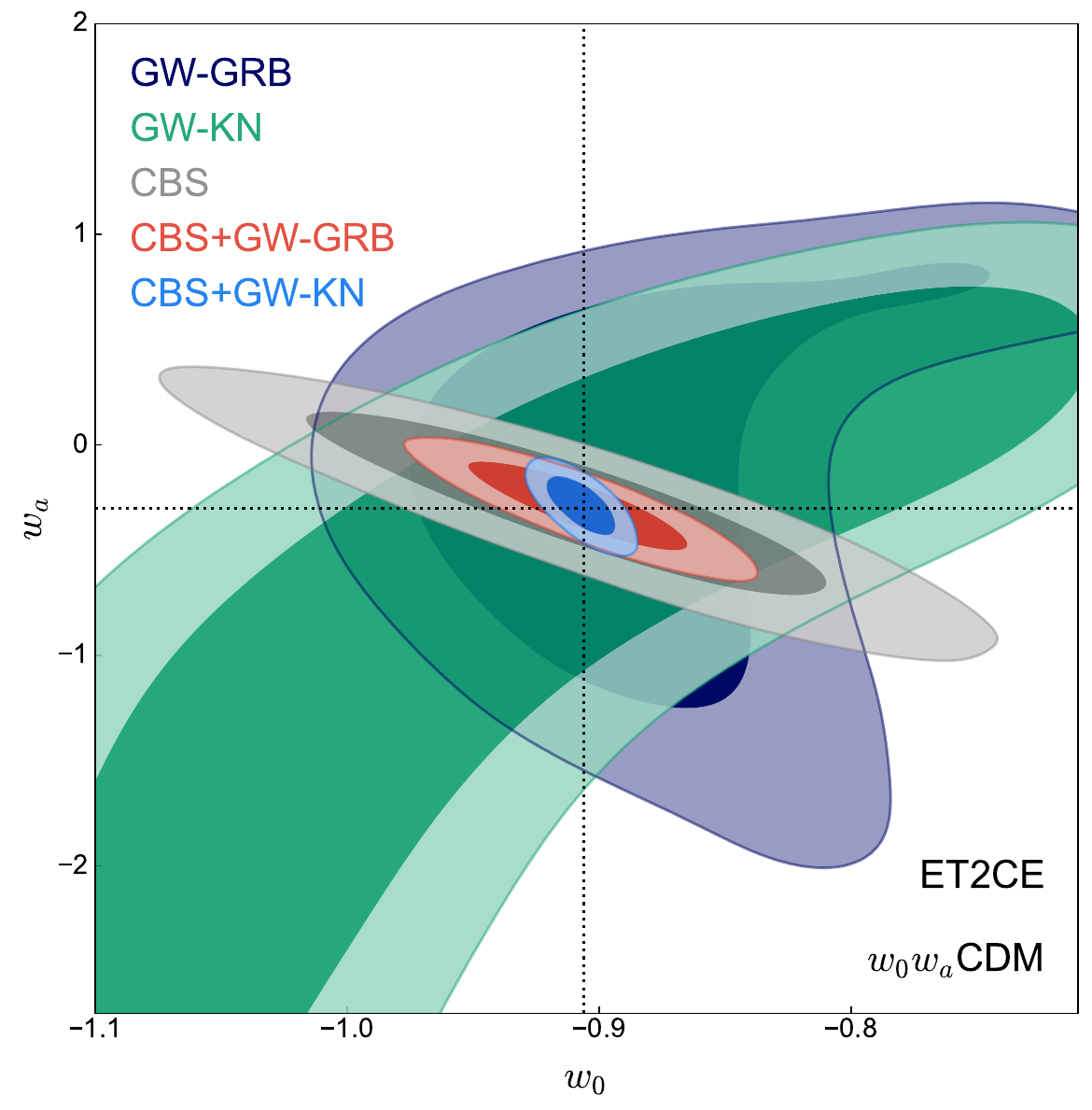}
	\caption{\label{fig14} Two-dimensional marginalized contours ($68.3\%$ and $95.4\%$ confidence level) using the GW-GRB, GW-KN, CBS, CBS+GW-GRB, and CBS+GW-KN data. The left panel shows the $w$--$\Omega_{\rm m}$ plane in the $w$CDM model and the right panel shows the $w_0$--$w_a$ plane in the $w_0w_a$CDM model. Here, the GW data are simulated based on the GW detection strategy of ET2CE, and the dotted lines indicate the fiducial values of cosmological parameters preset in the simulation.}
\end{figure*}

As mentioned above, Fig.~\ref{fig8} shows that GW multi-messenger observations under different EM detection scenarios have different redshift distributions, but they are all at redshifts below $\sim 4$. When combined with the early-universe information from the CMB and the complementary constraints from BAO and SN, their differing degeneracy orientations make the combination effective in breaking the cosmological parameter degeneracies. In this subsection, we analyze the constraint results from GW multi-messenger observations combined with CBS observations under different EM detection scenarios, aiming to demonstrate the capability of GW multi-messenger observations in breaking such degeneracies. Note that in the following discussions, we take ET2CE and LSST in the $i$ band as the representative GW and kilonova detection strategies. The constraint results are shown in Figs.~\ref{fig12}--\ref{fig14}, with details listed in Tables~\ref{tab7}--\ref{tab9}.

In Fig.~\ref{fig12}, we show the constraint results in the $\Lambda$CDM, $w$CDM, and $w_0w_a$CDM models using the CBS, GW-EM, and CBS+GW-EM data. As can be seen, the GW-EM data can tightly constrain the Hubble constant with precision ranging from $0.076\%$ to $0.034\%$. For the $\Lambda$CDM model, when using the GW-EM data, the constraints on $\Omega_{\rm m}$ and $H_0$ are both better than those from CBS. For the constraint on $w$ in the $w$CDM model, GW-EM provides a better constraint than that from CBS. In the case of the $w_0w_a$CDM model, GW-EM provides a tighter constraint on $w_0$ but a looser constraint on $w_a$ compared to CBS. Moreover, we can clearly see that the parameter degeneracy orientations of GW-EM and CBS differ significantly, and their combination effectively breaks cosmological parameter degeneracies.  With the addition of GW-EM to CBS, the constraint precision on cosmological parameters improves substantially. Specifically, for the $\Lambda$CDM model, CBS+GW-EM gives $\sigma(\Omega_{\rm m})=0.0010$ and $\sigma(H_0)=0.016$, which are $83.1\%$ and $96.2\%$ better than those from CBS. Furthermore, the constraint precision of $\Omega_{\rm m}$ and $H_0$ is $0.32\%$ and $0.024\%$, both exceeding the $1\%$ threshold for precision cosmology. For the $w$CDM and $w_0w_a$CDM models, combining GW-EM with CBS improves the constraints on $w$, $w_0$, and $w_a$ by $75.0\%$, $86.8\%$, and $69.0\%$, respectively. In particular, CBS+GW-EM achieves $\sigma(w) = 0.007$ in the $w$CDM model and $\sigma(w_0) = 0.009$ in the $w_0w_a$CDM model, with precision of $0.72\%$ and $0.99\%$. These results have already met the standard of precision cosmology.

In Fig.~\ref{fig13}, we present the constraint results in the $\Lambda$CDM, $w$CDM, and $w_0w_a$CDM models using the CBS+GW-GRB, CBS+GW-kilonova, and CBS+GW-EM data. We can see that CBS+GW-EM gives the best constraints on cosmological parameters, while CBS+GW-kilonova gives similar constraints to those from CBS+GW-EM. For instance, in the $\Lambda$CDM model, the constraints on $\Omega_{\rm m}$ and $H_0$ from CBS+GW-EM can be improved by $23.1\%$ and $74.6\%$ over those from CBS+GW-GRB. It is worth noting that, although GW-kilonova provides weaker constraints on $\Omega_{\rm m}$ and dark energy EoS parameters compared to GW-GRB in the $w$CDM and $w_0w_a$CDM models (as shown in Figs.~\ref{fig10} and~\ref{fig11}), their respective combinations with CBS demonstrate a reversal of this trend. To clarify this point, we present the constraint results using the GW-GRB, GW-kilonova, CBS, CBS+GW-GRB, and CBS+GW-kilonova data in the $w$--$\Omega_{\rm m}$ plane for the $w$CDM model and the $w_0$--$w_a$ plane for the $w_0w_a$CDM model in Fig.~\ref{fig14}. We find that although GW-kilonova provides weaker constraints on these parameters, it is more effective than GW-GRB in breaking the cosmological parameter degeneracies in terms of degeneracy directions generated by CBS. Therefore, when combining CBS with GW-kilonova, the constraints are better than those obtained by combining with GW-GRB.

\subsection{Constraint results from GW multi-messenger observations with different duty cycles}

Throughout this paper, in view of the high uncertainty of the duty cycle, we assume an optimistic scenario in which each GW detector operates with a 100\% duty cycle~\cite{Zhu:2021ram}. This represents an idealized case, as real detectors inevitably experience downtime due to maintenance, environmental disturbances, and unexpected technical issues, leading to lower effective duty cycles in practice. Therefore, we also consider the realistic scenario (assuming an 80\% duty cycle), and briefly discuss the impact of these two scenarios on cosmological analysis. Note that the following discussions are based on the GW detection strategy of ET2CE.

\begin{table*}[!htb]
	\caption{ Absolute (1$\sigma$) and relative errors of cosmological parameters in the $\Lambda$CDM, $w$CDM, and $w_0w_a$CDM models using GW-EM detections under different duty cycles. Here, the GW data are simulated based on the GW detection strategy of ET2CE, and $H_0$ is in units of $\rm km\ s^{-1}\ Mpc^{-1}$.}
	\label{tab10}
	\setlength{\tabcolsep}{2.75mm}
	\renewcommand{\arraystretch}{1.5}
	\begin{center}{\centerline{
				\begin{tabular}{c|c|c|c}
					\hline \hline
					\multirow{2}{*}{Model} &\multirow{2}{*}{Error} &\multirow{2}{*}{\shortstack{80\% duty cycle \\ (realistic scenario)}}   & \multirow{2}{*}{\shortstack{100\% duty cycle \\ (optimistic scenario)}} \\ &&\\
					\hline
					\multirow{4}{*}{$\Lambda$CDM}
					&$\sigma(\Omega_{\rm m})$       &$0.0025$   &$0.0023$      \\
					&$\sigma(H_0)$                            &$0.026$     &$0.023$          \\
					&$\ve(\Omega_{\rm m})$             &$0.79\%$ &$0.73\%$    \\
					&$\ve(H_0)$                                  &$0.039\%$ &$0.034\%$    \\  \hline
					\multirow{6}{*}{$w$CDM}
					&$\sigma(\Omega_{\rm m})$       &$0.0100$   &$0.0091$       \\
					&$\sigma(H_0)$                            &$0.036$     &$0.033$         \\
					&$\sigma(w_0)$                           &$0.023$       &$0.020$        \\
					&$\ve(\Omega_{\rm m})$             &$3.12\%$ &$2.83\%$     \\
					&$\ve(H_0)$                                  &$0.054\%$ &$0.049\%$    \\
					&$\ve(w_0)$                                  &$2.37\%$ &$2.06\%$   \\ \hline
					\multirow{7}{*}{$w_0w_a$CDM}
					&$\sigma(\Omega_{\rm m})$       &$0.0355$   &$0.0290$         \\
					&$\sigma(H_0)$                            &$0.059$     &$0.051$             \\
					&$\sigma(w_0)$                           &$0.031$      &$0.025$         \\
					&$\sigma(w_a)$                           &$0.43$         &$0.35$             \\
					&$\ve(\Omega_{\rm m})$              &$11.79\%$ &$9.32\%$   \\
					&$\ve(H_0)$                                  &$0.088\%$ &$0.076\%$   \\
					&$\ve(w_0)$                                  &$3.49\%$    &$2.78\%$     \\
					\hline
					\hline
		\end{tabular}}}
	\end{center}
\end{table*}

As shown in Table~\ref{tab10}, these two scenarios give similar constraint results, although the constraint results of the realistic scenario are slightly worse than those of the optimistic scenario. This indicates that our main conclusions remain robust even under the realistic scenario of detector operation, though further reductions in duty cycle could result in a somewhat reduced constraint capability.

\section{Conclusion}\label{sec5}

In this paper, we extend the previous work of Ref.~\cite{Han:2023exn} to explore the potential of multi-messenger GW standard sirens in the era of 3G GW detectors for constraining cosmological parameters. We conduct a comprehensive analysis of GW multi-messenger observations from BNS mergers using different EM detection scenarios, including GW-GRB, GW-kilonova, and their combined GW-EM observations. Two GW detection strategies are considered: the single ET and the ET2CE network. For the GRB observations, we select GECAM as a representative instrument due to its all-sky FoV, high sensitivity, and wide energy range. For the GW-kilonova observations, we consider three survey projects, namely WFST, LSST, and CSST, each of which employs three commonly used filters ($g$, $r$, $i$). Finally, we perform cosmological analyses using three typical models: the $\Lambda$CDM, $w$CDM, and $w_0w_a$CDM models.

We first predict the detection rates of future GW-GRB and GW-kilonova detections. We find that LSST in the $i$ band provides the highest number of detectable GW-kilonova events, with $\sim 4900$ events expected with redshifts below $\sim 0.4$ in a 10-year observation. In contrast, GW-GRB detections cover a much larger redshift range. Over a 10-year observation, the single ET is expected to detect 992 GW-GRB events with a redshift distribution below $\sim 2.4$, while the ET2CE increases the number of detectable GW-GRB events to 1455, extending the redshift range to $\sim 4$. We also find that at redshifts below $\sim 0.4$, EM counterparts are predominantly identified through kilonova detections, whereas at redshifts above $\sim 0.4$, GRBs play a dominant role in EM counterpart detections.

Then, we investigate the impact of GW multi-messenger observations under different EM detection scenarios on the constraints of cosmological parameters. For the GW-kilonova detections, we find that the $i$ band of LSST provides the best constraints, with similar constraints obtained from the $i$ band of CSST. Therefore, we choose LSST in the $i$ band as the representative survey project for GW-kilonova detections in our cosmological analysis. We find that GW-EM data can tightly constrain the Hubble constant with a precision ranging from $0.076\%$ to $0.034\%$. For the $\Lambda$CDM model, the constraints from GW-EM are stronger than those from GW-GRB and slightly better than those from GW-kilonova. For the $w$CDM and $w_0w_a$CDM models, although GW-kilonova provides weaker constraints on $\Omega_{\rm m}$ and the dark energy EoS parameters compared to GW-GRB detections, the combined GW-EM leads to improved constraints over both GW-GRB and GW-kilonova.

Finally, we analyze the constraint results from GW multi-messenger observations combined with CBS observations under different EM detection scenarios. We find that GW multi-messenger observations can effectively break the cosmological parameter degeneracies generated by traditional EM observations, and their combination will further improve the measurement precision of cosmological parameters. When combining GW-EM data with CBS, the constraint precision on the EoS parameters of dark energy $w$ in the $w$CDM model and $w_0$ in the $w_0w_a$CDM model can reach $0.72\%$ and $0.99\%$, which meet the standard of precision cosmology. We also find that, compared to GW-GRB, GW-kilonova is more effective at breaking the parameter degeneracies generated by CBS. Therefore, the combination of CBS and GW-kilonova provides better constraints than that of CBS and GW-GRB.

In the next few decades, the observations of multi-messenger GW standard sirens from 3G GW detectors and planned ground- and space-based telescope facilities can significantly enhance the constraints on cosmological parameters. With future upgrades of GW and EM facilities, we believe that multi-messenger standard siren cosmology will greatly facilitate the era of precision cosmology, playing a crucial role in achieving arbitration for the Hubble tension and providing insights into the expansion history of the universe.

\begin{acknowledgments}

This work was supported by the National Natural Science Foundation of China (Grants Nos. 12575049, 12533001, and 12473001), the National SKA Program of China (Grants Nos. 2022SKA0110200 and 2022SKA0110203), the China Manned Space Program (Grant No. CMS-CSST-2025-A02), and the 111 Project (Grant No. B16009).

\end{acknowledgments}

\bibliography{gw_em}

\end{document}